\documentclass[
 reprint,=
 amsmath,amssymb,
 aps,unsortedaddress]{revtex4-2}
\usepackage{xcolor}
\usepackage{graphicx}
\usepackage{dcolumn}
\usepackage{bm}

\newcommand{\be}{\begin{equation}}
\newcommand{\ee}{\end{equation}}

\usepackage{multirow}
\usepackage{color}
\usepackage{xcolor}
\usepackage{times}
\usepackage{amsmath,bm,amsfonts}
\usepackage{dcolumn}
\usepackage{latexsym}
\usepackage{hhline}
\usepackage{braket}
\usepackage{bbold}

\usepackage{adjustbox}
\usepackage{dsfont}
\usepackage{color}
\usepackage{xcolor}

\DeclareFontFamily{U}{mathx}{}
\DeclareFontShape{U}{mathx}{m}{n}{<-> mathx10}{}
\DeclareSymbolFont{mathx}{U}{mathx}{m}{n}
\DeclareMathAccent{\widehat}{0}{mathx}{"70}
\DeclareMathAccent{\widecheck}{0}{mathx}{"71}

\begin{document}

\title{Anyon Condensation Web and Multipartite Entanglement in 2D Modulated Gauge Theories}

\author{Guilherme Delfino}
\affiliation{Department of Physics, Boston University, MA, 02215, USA}
\author{Yizhi You}
\affiliation{Department of Physics, Northeastern University, MA, 02115, USA}

\date{\today}
\begin{abstract}

In this work, we introduce an anyon condensation web that interconnects a broad class of 2D finite gauge theories with multipolar conservation laws at a microscopic level. We refer to such theories as spatially modulated since their generators act non uniformly across the system and have a strong position dependence. We find that condensation of appropriate set of anyons triggers the emergence of additional spatially modulated symmetries, which has the general effect of increasing the number of super-selection anyon sectors. As explicit examples, we start with the rank-2 toric code model and implement various anyon condensation protocols, resulting in a range of 2D higher-rank theories, each with a distinct gauge structure. We also expand the scope of anyon condensation by introducing lattice defects into spatially modulated theories and demonstrate that these geometric defects can be viewed as effective anyon condensations along the branch cut. Furthermore, we introduce the Multipartite Entanglement Mutual Information measure as a diagnostic tool to differentiate among various distinct multipole conserving phases. A captivating observation is the UV sensitivity of the mutual information sourced from multipartite entanglement in such modulated gauge theories, which depends on the geometric cut and the system size, and exhibits periodic oscillations at large distances.

\end{abstract}

\maketitle

\section{Introduction}

Understanding the nature of entanglement in quantum field theory has led to important developments in the theoretical understanding and classification of quantum phases. In the past decades, zoology of quantum stabilizer codes \cite{kitaev03,wen03,shor1995scheme,steane1996error} has significantly contributed to the understanding of discrete gauge theories through exactly-solvable Hamiltonians \cite{atiyah1988topological,wen2015construction,wen2003quantum,witten1991quantization,kitaev2010topological}. Specifically, these stabilizer codes can be interpreted as an emergent gauge theory whose gauge structure is incorporated as the higher-form symmetries of the microscopic Hamiltonian \cite{wen2019emergent,kapustin2017higher,rayhaun2023higher}. Recently, the concept of \textit{fracton topological order} in three spatial dimensions, and higher, had drawn attention from both the high energy and condensed matter community \cite{Haah2011-ny,Vijay2015-jj,Chamon2005-fc,alicki2009thermalization,castelnovo2007entanglement,shirley2019universal,shirley2017fracton,shirley2018fractional,slagle2019symmetric,yoshida2015bosonic,hirono2022symmetry,you2020fractonic,you2020higher}, achieved as low-energy states of gapped subsystem charge and/or multipole moment conserving gauge theories \cite{gromov2019towards,you2019multipolar}. In two dimensions, gauging discrete multipole symmetries lead to rich symmetry enriched topological orders (SETs), where both translations and rotations act non-trivially on the anyon content \cite{oh2022rank,oh2023aspects,delfino22}. Alternatively, they can be viewed as twisted copies of usual discrete topological orders, where anyons change flavors when going through the periodic boundaries \cite{KLW0834, pace-wen}. For $\mathbb{Z}_N$ polynomial gauge theories, the mobility restriction in the quasiparticles content is only up to $\mathcal{O}(N)$ sites, due the  mod $N$ conservation of modulated quantities. A richer structure arises when gauging non-polynomial symmetries, e.g. \textit{exponential symmetries}, where complicated long operators are needed to move excitations arroud \cite{sala2022dynamics, delfino20232d}. The restricted mobility of quasiparticles can be naturally interpreted in terms of modulated charge conservation laws \cite{bulmash2018higgs,pretko2018fracton,pretko2017generalized,you2019fractonic,you2018symmetric,radzihovsky2020fractons,pace-wen,nguyen2020fracton}, often resulting in UV dependence of the ground state degeneracy on lattices with periodic boundary conditions \cite{oh2022rank,oh2023aspects,delfino22,delfino20232d}, a characteristic properties of topological phases  enriched by translations \cite{Kou2008_mutual}.

Although proposals have been made regarding the zoology of fracton stabilizer codes with distinct higher-rank gauge structures, a significant question still lingers: What is the interrelation between different modulated gauge theories? In the context of three-dimensional theories, the authors in Ref. \cite{aasen2020topological,song2023topological,Vijay2017-ey,Ma2017-qq, ma2018fracton} successfully bridge the gap between various types of gauge theories. It includes both the conventional 3D topological quantum field theories (TQFT) and 3D fracton gauge theories by adopting a unified defect network perspective. It suggests that diverse topological field theories and fracton gauge theories can be related to each other through  defect network approaches, which are achieved by imposing condensation of anyons on defect lines and planes.
Somehow related to these ideas, in this work our goal is to intricately intertwine a broad class of two-dimensional modulated gauge theories through the condensation of anyonic quasiparticles. 

We are interested in lattice models obtained from  gauging symmetries associated with modulated generators
\begin{eqnarray}\label{generator}
    G[f] = \sum_r f_r \, q_r
\end{eqnarray}
where $q_r$ corresponds to the charge density of quasiparticles at site $r = (x,y)$ and $f_r$ is a fixed integer lattice function that defines the modulated symmetry. For concreteness, $G[1]$ corresponds to the global charge, $G[x]$ and $G[y]$ are the $x$- and $y$-components of dipole momentum, and $G[x^2 - y^2]$ and $G[xy]$ are quadrupole momentum, etc. Although in this work we only focus on polynomial symmetries, the generators in Eq.~\eqref{generator} can be more exotic symmetries by choosing non-polynomial functions. As explicit examples, we mention exponential symmetries, with $f_r = a^{x+y}$ for some parameter $a$ (mod $N$); and subsystem symmetries, with $f_r = \delta_r(\Omega)$, 
\begin{eqnarray}
    \delta_r(\Omega) = \begin{cases}
        1, \quad \text{if}\quad r\in \Omega\nonumber\\
        0, \quad \text{otherwise}
    \end{cases}
\end{eqnarray}
for $\Omega$ a subset of lattice points where the symmetry operators have non-trivial support on. In 3D cubic lattices, gauging $G[\delta_r(\Omega)]$ for $\Omega$ corresponding to planes or fractal membranes can give rise to well known fracton codes, as the X-Cube and Haah's code \cite{x-cube}.

In this work, we strive to demonstrate that a variety of 2D discrete modulated gauge theories \cite{oh2022rank,oh2023aspects,delfino22,delfino20232d}, whose  symmetries are generated by Eq.~\eqref{generator}, associated to different polynomial functions $f_r$ can be connected to each other through an \textit{anyon condensation web}. Specifically, the condensation of a subset of quasiparticles confines dual excitations, that have nontrivial braiding statistics with them, thereby altering the underlying higher-rank gauge structure. This do not correspond to usual anyon condensation, as we modulate the condensing potential through space, explicitly breaking some of the lattice symmetries.  By meticulously selecting the type of anyon condensation, we can obtain a web of stabilizer models that connect various modulated gauge theories in two dimensions. 

Generally, we study phase transitions between phases associated with the ultra-infrared (IR) limit of gauge theories linked to two sets of generators: \( G[f_1], G[f_2], \ldots, G[f_n] \) and \( G[g_1], G[g_2], \ldots, G[g_m] \). Schematically, the gauge structure before and after the transition can be represented as:
\begin{eqnarray}\label{transition12}
    G[f_1]\oplus \ldots\oplus G[f_n] \rightarrow G[g_1]\oplus \ldots \oplus G[g_m],
\end{eqnarray}
where the arrow indicates the condensation of a set of appropriate anyons. Surprisingly, we find that the resulting condensated phase sometimes have more global anyonic super-selection sectors than the uncondensed phase. This is captured by the total quantum dimension $\mathcal D = \sqrt{\sum_a d_a}$ which corresponds to the square root of the number of independent anyons in the theory since $d_a =1 $ for Abelian anyons. Generically, in the transitions represented in Eq. \eqref{transition12}, we find that
\begin{eqnarray}
\mathcal{D}_{\text{uncondensed}}\leq \mathcal{D}_{\text{condensed}}.
    \end{eqnarray}
The fact that more anyons can emerge after the transition is a direct consequence of their non-trivial transformations under lattice symmetries. This is to be contrasted to usual Abelian anyon condensation transitions, where the number of super-selection sectors is smaller in the condensed phase, since some of the anyons have condensed and some others have confined \cite{Burnell_2018}. As we illustrate in the main text, this follows from the modification of the Gauss law and magnetic flux in the condensed, allowing for emergent additional conserved quantities.

\begin{figure}[h!]
    \centering
\includegraphics[scale=0.25]{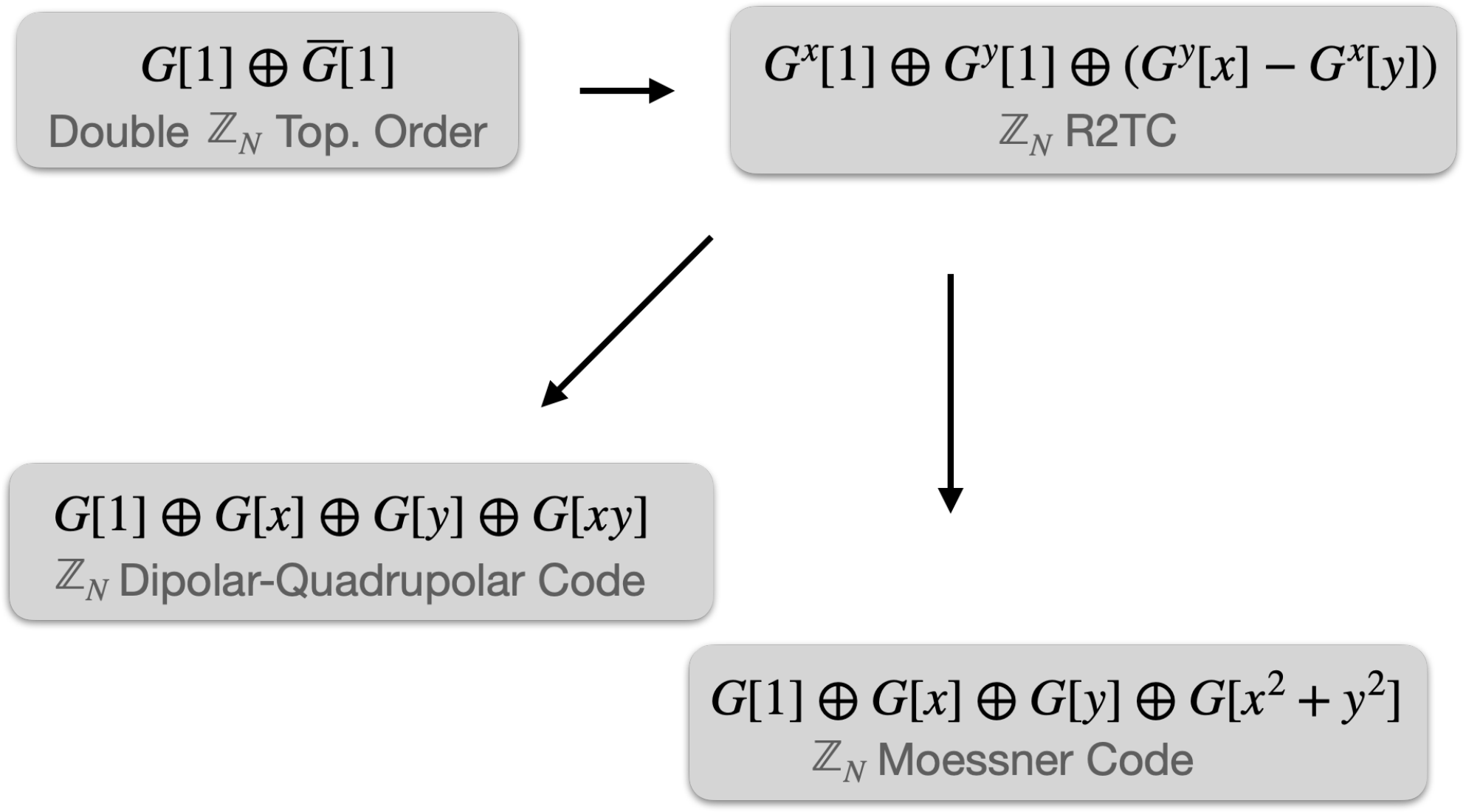}
    \caption{A small section of an anyon condensation web. The arrows correspond to phase transitions implemented through anyon condensation.}
    \label{web}
\end{figure}

In a nutshell, in the first part of this work we have explored a small section of a large web of theories that can be connected to each other through phase transitions implemented by anyon condensation, as illustrated in Fig. \ref{web}. The theories depicted in gray boxes are well known in the literature, and the arrows indicate the condensation of an appropriate set of anyons when transitioning between the SETs. In Fig. \ref{summary} we show a summary table with all the main properties of such different topological phases. 

Additionally, in a slightly different context, we study lattice defects, such as dislocations or disclinations, as mechanisms for interchanging different charge and flux sectors. This occurs due to the fact that distinct topological sectors in spatially modulated gauge theories undergo nontrivial permutations under translation and rotation. We also posit that geometric defects in modulated gauge theories can be perceived as branch lines subject to the condensation of appropriate anyons \cite{you2019non,manoj2020screw,aitchison2023no}. We then proceed to study the implications on the gauge structure of the theory when such condensation defects are present. 

We believe our findings might have significant implications for both the theoretical and eventual experimental understanding of 2D modulated gauge theories.
The anyon condensation transitions we consider provide a novel playground for exploring unconventional phase transitions beyond the Landau-Ginzburg-Wilson (LGW) paradigm within systems possessing UV/IR scale mixing. From an alternative perspective, anyon condensation can also be implemented through partial measurements on a quantum many-body wave function \cite{verresen2021efficiently,lu2022measurement,choi2021emergent,cotler2023emergent}. Explicitly, by measuring some qubits in the ground state wave function of a higher-rank stabilizer code, the post-measurement state—after error correction—is the ground state of another type of modulated gauge theory. In this context, partial measurement procedures play the role of implementing anyon condensation on the quantum many-body wave function, suggesting that various spatially modulated phases are connected to each other through partial measurements.

Finally, we also propose the \textit{multi-partite entanglement mutual information} \cite{jian2015long} as a strategy to differentiate between distinct 2D modulated gauge theories ground states. While entanglement entropy has been studied in the context of spatially-modulated phases \cite{Ebisu2023entanglement,Kim2023defects}, here the idea is that such modulated theories are characterized by their unique gauge structures and holonomies \cite{han2018_entanglement, Shi2018_entanglement, Dua2020_entanglement}. 
By performing different geometric cuts on the wave function the long-range correlation between Wilson operators should be reflected in the multipartite entanglement.
Effectively, it captures the underlying spatially modulated holonomy structures defined in geometric sectors. An important factor in multi-partite entanglement mutual information in spatially modulated gauge theories is that it is not only influenced by the system size but also by the distance (with periodic oscillations) and orientation of the cuts, incorporating a strong geometric dependence and UV/IR mixing.

\begin{widetext}
\begin{center}
    \begin{figure}
        \centering
\includegraphics[scale=0.40]{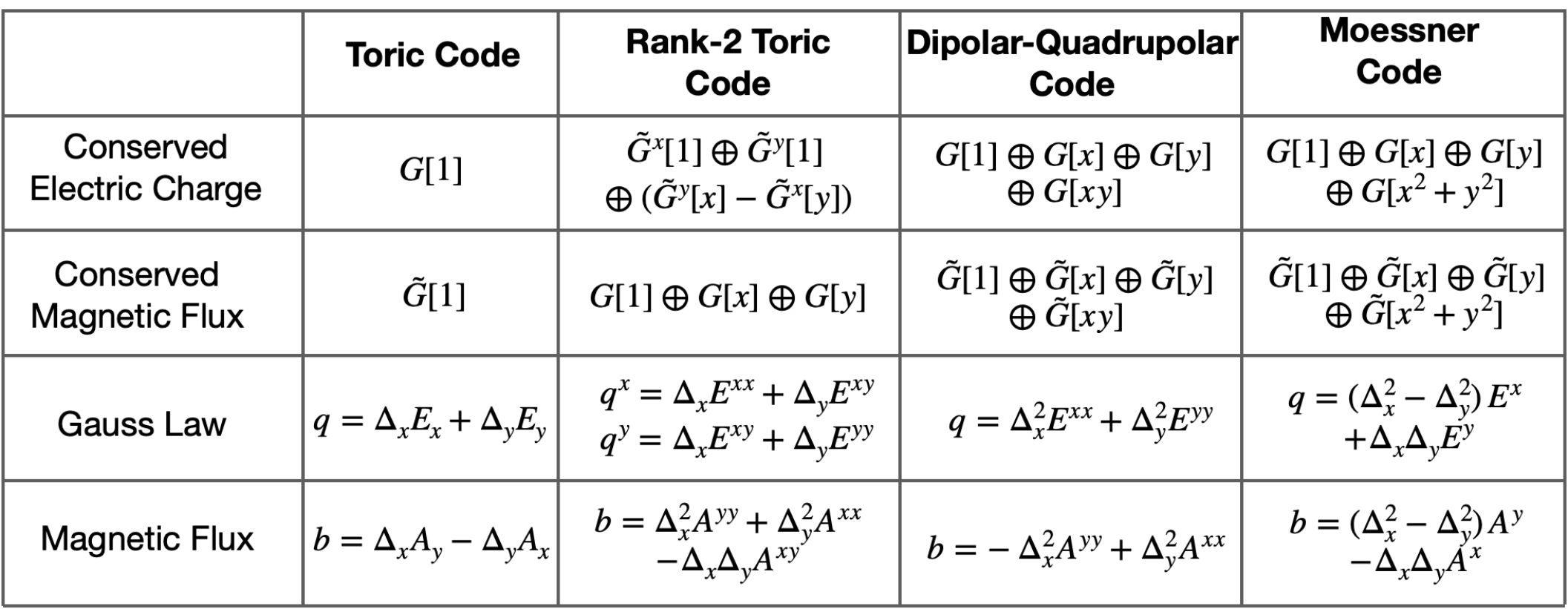}
        \caption{Summary table of properties associated to several two-dimensional spatially modulated $\mathbb Z_N$ gauge theories discussed in this work. The table shows the conserved modulated electric and magnetic quantities $G[f]$ and $\tilde G[\tilde f]$, respectively. It also shows the local Gauss law and the gauge-invariant magnetic flux for each modulated gauge theory.}
        \label{summary}
    \end{figure}
    \end{center}
\end{widetext}

\section{Remarks in Spatially Modulated Gauge Theories}

We often refer to spatially modulated gauge theories through the generators $G[f]$ of their gauged $\mathbb Z_N$ symmetries. When gauging the global $\mathbb Z_N$ symmetry associated with $G[f]$, we introduce gauge fields $A_a$ and their canonical conjugated electric fields $E_a$, for some suitable set of indices $a$. The gauge transformations $A_a\rightarrow A_a + \Delta_a \alpha$ are generated by the Gauss-Law, schematically represented as
\begin{eqnarray}\label{gausslaw}
    q_r = \Delta_a E_a,
\end{eqnarray}
where $\Delta_a$ is a generalized difference operator that contains information about the lattice function $f_r$. It is defined through the annihilation property $\Tilde \Delta_a f_r = 0$, where
\begin{eqnarray}
    \sum_r g_r \, \Delta_a h_r = \sum_r h_r\, \Tilde \Delta_ag_r,
\end{eqnarray}
for any lattice functions $g_r$ and $h_r$. 
The above definition can be thought of as an integration by parts, defining a generalized notion of the Leibniz rule.
Fixed a given function $f_r$, the annihilation property $\Tilde \Delta_a f_r = 0$ does not completely fix the derivative operator $\Delta_a$. A simple example is to consider a polynomial function $f_r$ of degree $n$, which is annihilated by any $m-$th order derivative. If $m>n+1$, however,  the theory conserves unwanted extra quantities $G[g]$, with $g$ any polynomial of degree $m-1$. We thus want to choose $\Delta_a$ carefully such that only $G[f]$ is conserved. 

One can also define a gauge invariant magnetic flux, schematically expressed as
\begin{eqnarray}\label{magneticflux}
    b_r = \widecheck\Delta_a A_a,
\end{eqnarray}
where $\widecheck\Delta_a$ is defined such that $    \widecheck\Delta_a\Delta_a \alpha = 0$
for any lattice function $\alpha$. This condition enforces that $b_r$ is gauge invariant. The choices of $\Delta_a$ and $\widecheck \Delta_a$ completely specifies the gauge theory.

The Noether's charges associated to $G[f]$ are still conserved in the gauge theory through its Gauss law. From inspection, one can see that $G[f]$ is a conserved quantity in the whole lattice
\begin{eqnarray}\label{constrain}
    G[f] &=& \sum_r f_r\, \Delta_a E_a\nonumber\\
    &=& \sum_r (\tilde \Delta_a f_r)\,  E_a = 0, 
\end{eqnarray}
which follows from definition $\tilde \Delta_a f_r = 0$. In general, a similar constraint for the magnetic fluxes also exist
\begin{eqnarray}\label{constrain_mag}
    \tilde G[\tilde f] &=& \sum_r \tilde f_r\,  \widecheck\Delta_a A_a\nonumber\\
    &=& \sum_r ( \tilde{\widecheck\Delta}_a \tilde f_r)\,  A_a = 0, 
\end{eqnarray}
for dual functions $\tilde f_r$ that are annihilated by $\tilde{\widecheck\Delta}_a$.
The fact that $G[f]$ vanishes (Eq. \eqref{constrain}) for any gauge invariant state in the Hilbert space implies in constrained dynamics for quasi-particles. The excitations can only move in ways such that its dynamics respects Eq. \eqref{constrain}, signaling for the presence of fracton-like behavior.

For convenience, we introduce the $\mathbb Z_N$ notation $X_a= e^{i2\pi E_{a}/N}$ and $Z_a= e^{i A_a}$, which obey the clock algebra $X_a Z_b = \omega^{\delta_{ab}} Z_b X_a$, with $\omega = e^{2\pi i/N}$. It is also convenient to introduce $\mathbb Z_N$ charge $\mathcal Q_r$ and magnetic fluxes $\mathcal B_r$ operators,
\begin{eqnarray}
    \mathcal Q_r &=& e^{\frac{2\pi i q_r}{N}} =  \prod_a X_a^{\Delta_a}\nonumber\\
    \mathcal B_r &=& e^{\frac{i b_r}{N}} =  \prod_a Z_a^{ \widecheck\Delta_a}.
\end{eqnarray}
Here, $\Delta_a$ and $\widecheck\Delta_a$ should be understood in terms of the coefficients that accompany the terms of $E_a$ and $A_a$ in the Gauss law and magnetic fluxes in Eq. \eqref{gausslaw} and \eqref{magneticflux}.

In this notation the conservation laws become constraints for the allowed charges and fluxes eigenvalues
\begin{eqnarray}
    \prod_r \mathcal Q_r^{f_r} = \mathds{1}\nonumber\\
    \prod_r  \mathcal B_r^{\tilde f_r} = \mathds{1}.\label{con123}
\end{eqnarray}
As we discuss later through some examples, under periodic boundary conditions the $\mathbb Z_N$ gauge theory can be reduced to a $\mathbb Z_k$ gauge theory, where $k$ might depend on the system sizes. This follows from the imposition that the constraints in \eqref{con123} are well defined, implying in twisted boundary conditions for the gauge fields.

The higher form symmetries of the $\mathbb Z_N$ (or $\mathbb Z_k)$  modulated gauge theories can be made explicit by taking the product $\mathcal Q_r$ and $\mathcal B_r$ in a finite region $\mathcal A$. Let $\partial \mathcal A$ be the boundary $\mathcal{A}$, then we have that the product
\begin{eqnarray}
    \prod_{r\in \mathcal A} \mathcal Q_r^{f_r} = W^f(\partial \mathcal A)\nonumber\\
    \prod_r \mathcal  B_{\mathcal A}^{\tilde f_r} = V^{\tilde f}(\partial \mathcal A),
\end{eqnarray}
reduces to string operators at the boundary $\partial{\mathcal A}$. Such closed line operators correspond to gauge invariant Wilson and t'Hooft loops, which allow us to study excitations in the theory as well as their mobility properties. For this, we consider open strings $W^f$ and $V^{\tilde f}$, whose general effect is to excite electric charges and magnetic fluxes at their endpoints. The constrained mobility of anyonic excitations, frequently present in modulated gauge theories, are incorporated in the rigidity of such strings. 

As a final comment, instead of dealing with a constrained Hilbert space and identifying physical states as gauge invariant ones $(\mathcal Q_r-1)\ket{\psi} = 0$, we instead choose to enforce it energetically. This is convenient as we can interpret non-gauge invariant states as charge excitations in the spectrum of the theory. For this, we add the $\mathbb Z_N$ Gauss law and magnetic fluxes operators directly into the lattice Hamiltonian
\begin{eqnarray}\label{ham_gen}
     H = -\sum_r \mathcal Q_r - \sum_r \mathcal B_r + \text{h.c.}.
\end{eqnarray}
By construction, the model defined above is exactly solvable, as every term commutes with each other
\begin{eqnarray}
     \mathcal B_r \mathcal Q_r &=& \prod_a \, \omega^{- \widecheck\Delta_a \Delta_a}\mathcal  Q_r \mathcal B_r\nonumber\\
     &=& \mathcal Q_r \mathcal B_r,
\end{eqnarray}
where we used that $\widecheck\Delta_a \Delta_a = 0$.

\section{Review of R2TC from anyon condensation of toric code models}
\label{sec:r2tc}

\begin{figure}[tb]
\includegraphics[width=0.50\textwidth]{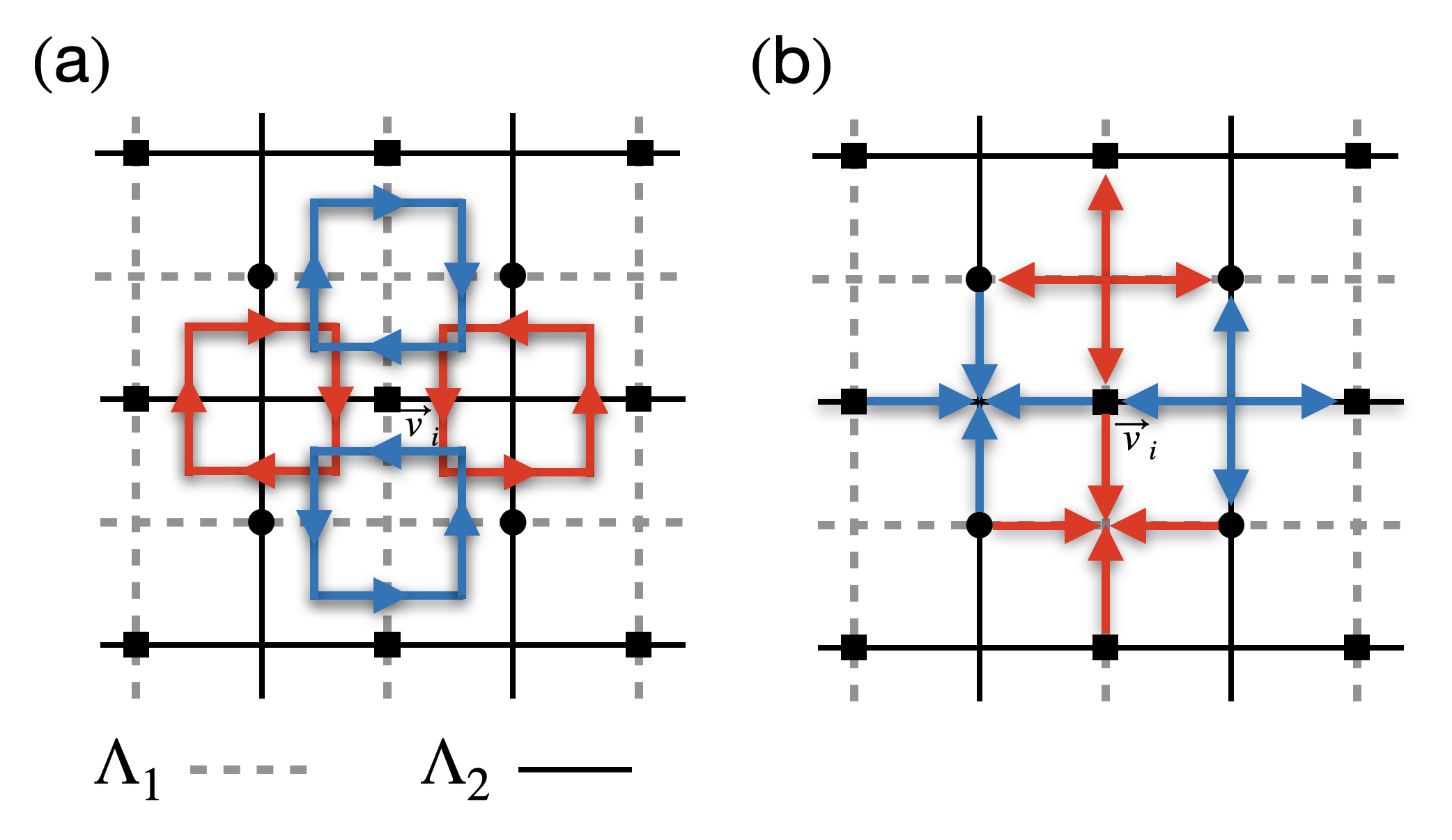}
\caption{
Illustration for (a) condensed magnetic flux $B$ and (b) condensed Gauss's law $G$ defined in Eqs. (\ref{r2tc}) on two interpenetrating square lattices.
The sublattice $1$ is shown by dashed lines and the sublattice $2$ is shown by the solid lines. The condensation of the fields takes place on the dotted links.
(a) The red and blue arrows represent the fields $A_{r,1}^a$ and $A_{r,2}^a$ ($a = x,y$), respectively, where the positive directions are right and up. (b) The red and blue arrows represent the fields $E_{r,1}^a$ and $E_{r,2}^a$ ($a = x,y$), respectively.
}
\label{fig:quadrupole}
\end{figure}

To set the stage, we briefly review the anyon condensate protocol introduced in Ref.~\cite{oh2023aspects}, where rank-2 gauge theory can emerge through anyon condensation from two copies of the rank-1 gauge theory. The details of this anyon condensate were elucidated in Ref.~\cite{oh2023aspects}; we reiterate them here as a prelude to introducing the general anyon condensate procedure later.
Let us begin with two sets of $\mathbb{Z}_N$ gauge theories living on the interpenetrating square lattices denoted $\Lambda_1$ (dashed lines) and $\Lambda_2$ (solid lines) as in Fig. \ref{fig:quadrupole}, with lattice vectors in the $x$- and $y$-directions given by $e_1$ and $e_2$. Each square lattice has gauge degrees of freedom $(A^\mu_{a,r} , E^\mu_{a,r})$ residing at the $\mu=x,y$-oriented links of the respective square sublattice labeled by $a=1,2$, at site $r$. They satisfy the canonical commutation $[A_{a,r}^\mu , E_{a',r'}^{\mu'} ] = i \delta_{aa'} \delta_{\mu \mu'}\delta_{r,r'}$. 
Each square lattice hosts a deconfined $\mathbb{Z}_N$ gauge theory as,
\begin{align} 
q_{a} = ({\bm \nabla} \cdot {\bm E}_a) ,
~~~ b_{a} = ({\bm \nabla} \times {\bm A}_a )
, \label{eq:2.1}\end{align} 
In the above, the charges $q_{a}$ can assume any integer value mod $N$ and the fluxes $b_{a}={2\pi k}/{N}$ with $(k=1,2,..N)$. Also, $\bm \nabla = (\Delta_1, \Delta_2)$ is the two-dimensional lattice derivative nabla operator, with
\begin{eqnarray}
    \Delta_i f_r \equiv f_{r+e_i} - f_r.
\end{eqnarray}
To implement the anyon condensation, we add a strong onsite interaction term $\cos[\frac{2\pi}{N} (E^x_{1,r+e_y}-E^y_{2,r+e_x})]$ illustrated in Fig. \ref{fig:quadrupole}. The local Hilbert space is projected to the subspace as:
\begin{align}
E^x_{1,r+e_y} = E^y_{2,r+e_x} \quad (\text{mod } N) \label{eq:constraint}. 
\end{align}
The projection operator can be viewed as an anyon condensation process that proliferates an anyon-bound state of a flux dipole of layer 1 oriented in the $x$-direction with a flux dipole of layer 2 oriented in the $y$-direction. Such procedure, in turn, confines operators $A^x_{1,r+e_y}, A^y_{2,r+e_x}$ due to their nontrivial mutual statistics. As a result, the magnetic field operator is no longer well defined and the leading-order gauge invariant operator that commutes with the constraint is,
\begin{align}
      b_r = \Delta_x ({\bm \nabla} \times {\bm A}_{1,r}) - \Delta_y ({\bm \nabla} \times {\bm A}_{2,r} )
 \label{eq:cond-b}
\end{align}
Such a term involves only the symmetric combination $A^x_{1} + A^y_{2}$, which one can explicitly verify its commuting property with the constraint $( E^x_{1} -  E^y_{2} ) =0$ mod $N$. 

To characterize the gauge theory after anyon condensation, we introduce a convenient notation
\begin{align} 
(E_{2,r}^x, \, E_{1,r}^y , \, E_{1,r+e_y}^x = E_{2,r+e_x}^y ) \rightarrow (E_r^{xx},\,  E_r^{yy}, \, E_r^{xy} ) . 
\end{align} 
Likewise,
\begin{align}
(A_{2,r}^x , \, A_{1,r}^y ,\,  A_{1,r+e_y}^x + A_{2,r+e_x}^y ) \rightarrow (A_r^{xx}, \, A_r^{yy},\,  A_r^{xy})
\end{align} 
which makes the relation to a rank-2 gauge theory $(A_r^a, E_r^a)$ ($a=xx,xy,yy$) more explicitly. From definition, the fields obey the canonical relation $[A_r^a, E_{r'}^b] = i \delta_{rr'} \delta_{ab}$ and transform under gauge as
\begin{align}\label{gauge_trans}
 &A^{xx} \rightarrow A^{xx} +\Delta_xf_1, \nonumber\\
 &A^{yy} \rightarrow A^{yy} +\Delta_yf_2, \nonumber\\
 &A^{xy} \rightarrow A^{xy} +\Delta_xf_2+\Delta_yf_1.
\end{align}

The resulting gauge theory is defined by the following Gauss laws and magnetic flux
\begin{align}
&q^x =  \Delta_x E^{xx} +\Delta_y E^{xy}\nonumber\\
&q^y = \Delta_x E^{xy} +\Delta_y E^{yy}\nonumber\\
&b = \Delta^2_x A^{yy}+\Delta^2_y A^{xx}-\Delta_x \Delta_y A^{xy},
\label{r2tc}
\end{align} 
where $q^x$ and $q^y$ are the gauge transformation generators associated to $f_1$ and $f_2$ in Eq. \eqref{gauge_trans}.
In the presence of periodic boundary conditions, this theory possesses a ground-state degeneracy that is sensitive to the linear system sizes $L_x$ and $L_y$ 
\begin{eqnarray}\label{gsd_r2tc}
    GSD = N^3\gcd(N,L_x) \gcd(N, L_y) \gcd(N, L_x, L_y).
\end{eqnarray}

The gauge theory specified in Eq.~\eqref{r2tc} corresponds to the rank-2 toric code (R2TC) model, proposed in Ref. \cite{oh2022rank} and corresponds to a discrete $\mathbb{Z}_N$ rank 2 vector tensor gauge theory. Such a theory has vector-like charges $q_x$ and $q_y$, which can be show to conserve the modulated quantities ,
\begin{eqnarray}\label{vector}
     G^x[1] = \sum_r q^x , \quad G^y[1] = \sum_r q^y, \nonumber\\
     \text{and}\quad G^y[x] - G^x[y] = \sum_r \left(x\, q^y - y\, q^x\right),
\end{eqnarray}
which follow from the Gauss law in Eq. \eqref{r2tc}.
 Similarly, one can explicitly check that the magnetic fluxes conserve
\begin{eqnarray}
    \tilde G[1]\oplus \tilde G[x]\oplus \tilde G[y],
\end{eqnarray}
where $\tilde G[\tilde f]=\sum_r \tilde f_r\, b_r$, as summarized in Fig.~\ref{summary}. Because of this asymmetry in the electric and magnetic conserved quantities, the electric and magnetic anyons of the R2TC obey different mobility restriction rules. Schematically, we represent the phase transition we discussed in the above as
\begin{eqnarray}
    G[1]\oplus \overline G[1]\rightarrow G^x[1]\oplus G^y[1]\oplus (G^y[x]-G^x[y]),
\end{eqnarray}
which encodes the Gauss laws of both theories before and after the transition.  
These results were originally studied in Ref. \cite{oh2022rank,oh22b,bulmash2018higgs}. 
In particular, Ref.~\cite{bulmash2018higgs} pointed out that when choosing \(N=2\), the \textit{vector rank-2 tensor gauge theory} is reminiscent of three copies of the toric code model (denoted as \(\mathbb Z_2^3\) gauge theory), with three flavors of ($\mathbb Z_2$) charge and flux. Notably, the three charges (fluxes) can be permuted by lattice translations/rotations, manifesting as a spatial symmetry-enriched topological phase where different anyons are intertwined with each other through crystalline symmetries. This novel interplay between anyon excitation and spatial symmetry is a signature of higher-rank gauge theory in 2D as well as fracton in 3D. Ref.~\cite{pace-wen} delineated that 2D \(
\mathbb Z_N\) higher-rank gauge theory can be treated as a spatial symmetry-enriched topological order as spatial symmetry permutes different anyon types. Likewise, Ref.~\cite{williamson19} suggests that 3D fractonic matter can naturally emerge from a 3D SPT phase with global U(1) and translational symmetries after we gauge the global U(1) symmetry. In the SPT state, the global symmetry quantum numbers of excited quasiparticles depend on their positions in a nontrivial way. Thus, after gauging U(1) symmetry, the resultant gauge charge in 3D exhibits restricted mobility as fractons.

\section{Anyon Condensation Web for 2D Modulated Theories}
\label{condensation}

From now on, we will consider the R2TC model as a starting point for our studies and demonstrate that a variety of 2D spatially modulated gauge theories can be derived from different anyon condensation schemes. The key concept is that in spatially modulated gauge theories, the gauge structure and conservation laws are associated with generalized higher-form symmetries, and their ground state degeneracy results from the spontaneous breaking of such symmetries. A specific type of anyon condensation 
necessarily confines all other excitations that have nontrivial mutual statistical interactions with them, thus altering the gauge structure and engendering a new, distinct, ordered phase. 

\subsection{A Route to 
Dipolar-Quadrupolar code}\label{sec:chamon}

In this section, we aim to establish a connection between the R2TC theory as represented in Eq. \eqref{r2tc}, and the Dipolar-Quadrupolar code introduced in Ref. \cite{delfino22}, and also studied in \cite{Ebisu2023,bulmash2018higgs}, via charge vector condensation. The operator $e^{i A^{xy}}$ generates a pair of vector dipole moments for both $q^x$ and $q^y$. Assume we condense these dipole moments by introducing a strong Higgs term that favors the constraint $A^{xy}=0$ ($ \text{mod}\, 2\pi$), inherently ruling out operators that do not commute with it, such as $E^{xy}$. Consequently, the vector charge $(q_x,q_y)$, as specified in Eq. \eqref{r2tc}, is no longer well-defined after Higgsing. In its place, a new Gauss law arises,
\begin{align}
  q \equiv  \Delta_x q^y- \Delta_x q^x=\Delta^2_x E^{xx} -\Delta^2_y E^{yy},
\end{align}
which commutes with the $A^{xy}=0$  ($ \text{mod}\, 2\pi$) constraint.
Accordingly, the magnetic flux operator after Higgsing yields,
\begin{align}
&b = \Delta^2_x A^{yy}+\Delta^2_y A^{xx}.
\end{align} 

\begin{figure}
    \centering
    \includegraphics[scale=0.4]{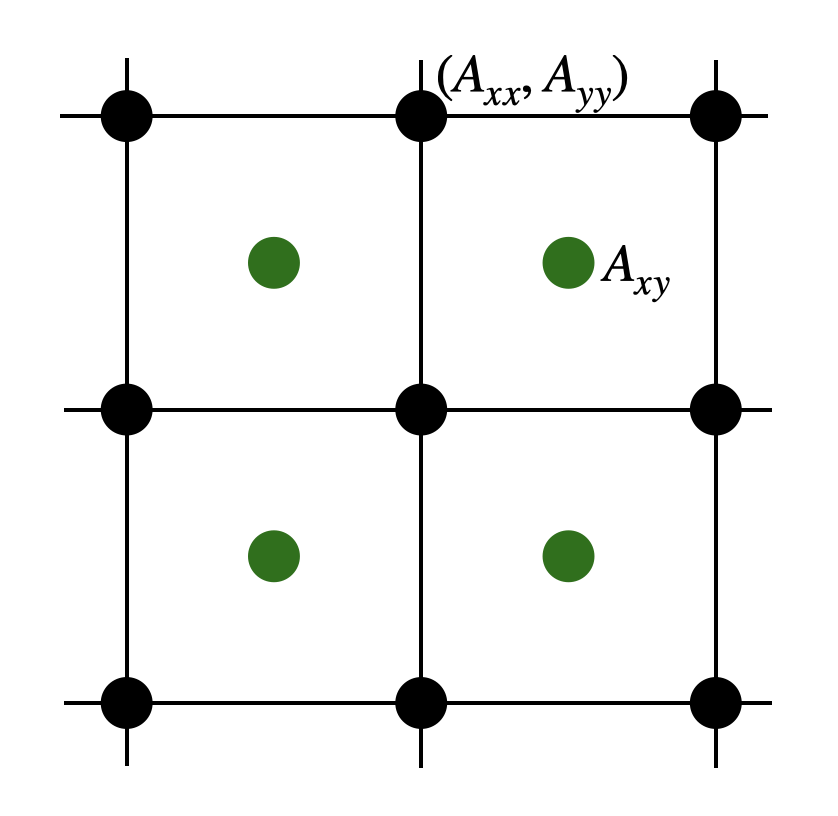}
    \caption{Gauge fields at vertices and face centers in Dipolar-Quadrupolar code on a square lattice.}
    \label{gauge_fields_DC}
\end{figure}

To maintain consistency with the notation used for the Dipolar-Quadrupolar code in Ref.~\cite{delfino22}, we adjust the labels for the electric field and gauge potential as follows:
\begin{align}
 &E^{yy} \rightarrow -E^{yy},  \quad 
 &A^{yy} \rightarrow -A^{yy}, 
\end{align} 
so the corresponding gauge theory can be rephrased as,
\begin{align}
&q=\Delta^2_x E^{xx} +\Delta^2_y E^{yy}\nonumber\\
&b = -\Delta^2_x A^{yy}+\Delta^2_y A^{xx}.
\label{precha}
\end{align} 
The two gauge potential follows a gauge transformation as,
\begin{align}
 &A^{xx} \rightarrow A^{xx} +\Delta^2_x f, \nonumber\\
 &A^{yy} \rightarrow A^{yy} +\Delta^2_y f.
\end{align} 

Before we delve deeper, it is essential to emphasize the gauge structure of this theory. Unlike the traditional lattice gauge theory — which assigns different gauge potentials to individual links—this theory distinctively places both gauge potentials $A^{xx},A^{yy}$ on the same lattice sites. The gauge theory detailed in Eq. \eqref{precha} represents a generalized electromagnetism that preserves charge $G[1]$, dipole moments $G[x], G[y]$, and quadrupole moment component $G[xy]$. The self-dual structure of the theory in Eq. \eqref{precha} indicates that magnetic flux shares a similar multipolar conservation law.

\begin{figure}
    \centering
    \includegraphics[scale=0.26]{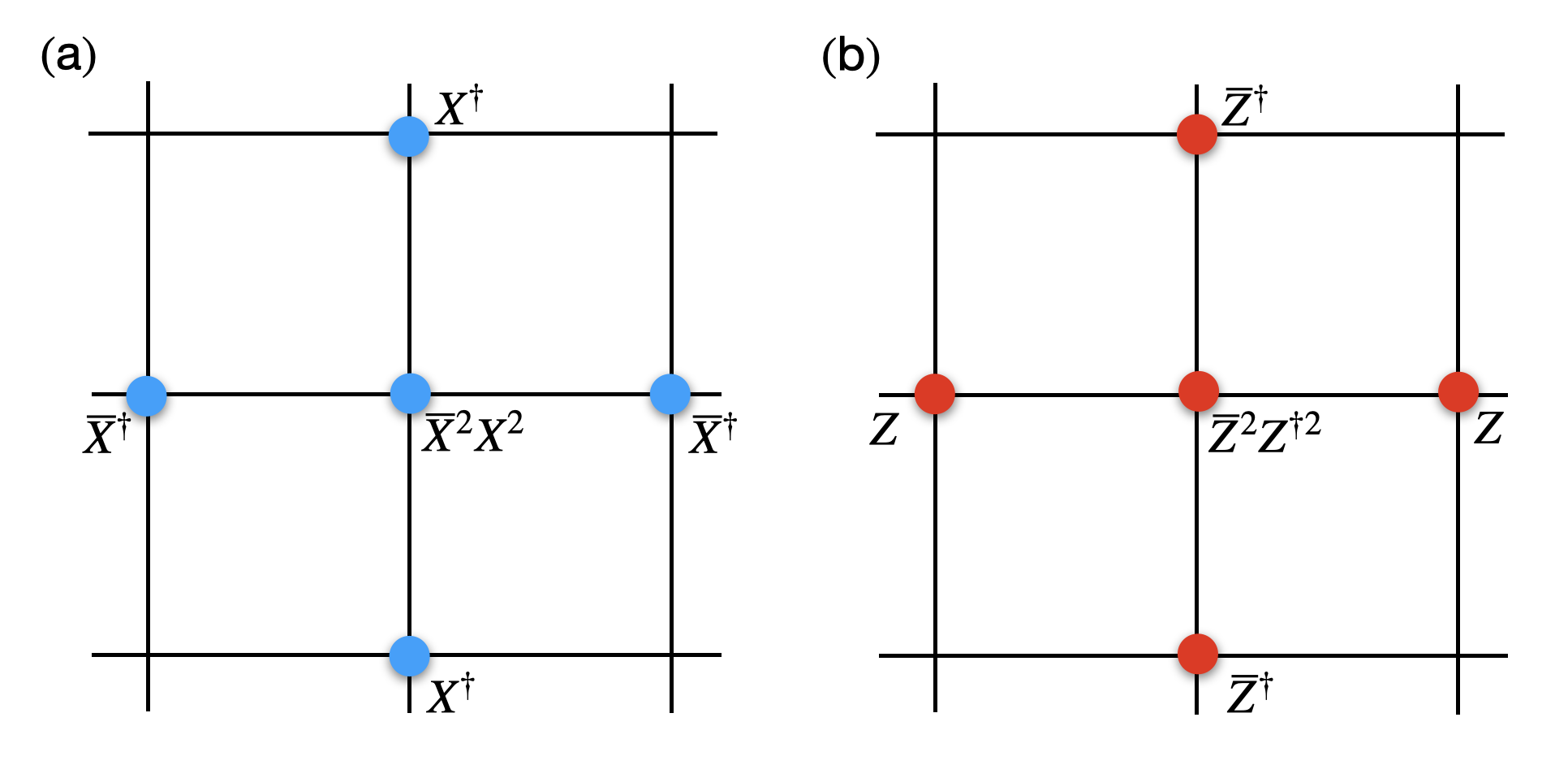}
    \caption{Discrete $\mathbb{Z}_N$ charge and flux operators in the doubled version of Dipolar-Quadrupolar code.}
    \label{Doubled_Delfino-Chamon}
\end{figure}

To express the gauge theory in Eq. \eqref{precha} in terms of a stabilizer code, we parameterize the field components as two sets of $\mathbb{Z}_N$ Pauli operators: $e^{i\frac{2\pi}{N}E^{yy}}=X, e^{iA^{yy}}=Z,e^{i\frac{2\pi}{N}E^{xx}}=\bar{X},$ and $e^{iA^{xx}}=\bar{Z}$. Under this notation, the gauge theory can be expressed in terms of a CSS-type stabilizer code,
\begin{eqnarray}\label{model1}
    H=-\sum_r\mathcal{Q}_r - \sum_r \mathcal{B}_r+\text{h.c.} 
\end{eqnarray}
with $\mathcal{Q}_r$ and $\mathcal{B}_r$ being the charge and flux operators, defined as
\begin{widetext}
\begin{eqnarray}
\mathcal{Q}_r  &=& \bar{X}^{\dagger}_{r-e_x} \, \bar{X}^{\dagger}_{r+e_x} \, \bar{X}_{r}^{2} \, X_{r}^{2} \,X^{\dagger}_{r+e_y} \, X^{\dagger}_{r-e_y}, \nonumber\\ \text{and}\quad \mathcal{B}_r  &=& Z_{r-e_x} \, Z_{r+e_x} \, Z_{r}^{-2} \, \bar{Z}_{r}^{2} \,\bar{Z}^{\dagger}_{r+e_y} \, \bar{Z}^{\dagger}_{r-e_y},
\label{css}
\end{eqnarray}
\end{widetext}
as illustrated in Fig. \ref{Doubled_Delfino-Chamon}.

  The commutative nature of these two terms contributes to the exact solvability of the ground state, generating a wave function characterized by patterns exhibiting zero flux and charge. It is worth mentioning that we added the Gauss law and magnetic flux terms directly into the Hamiltonian \eqref{model1}. In this sense, we are enforcing gauge invariance $\mathcal Q_r =1$ energetically, in the lowest energy states and do not impose further constraints in the Hilbert space.

As mentioned before, the model in Eq. \eqref{css} presents $\mathbb{Z}_N$ dipole,
and off-diagonal quadruple moment conservation and, as a consequence, restricts the mobility of excitations. We can see this anyon condensate transition as
\begin{eqnarray}\label{transition_delfino}
    G^x[1]\oplus G^y[1]\oplus (G^y[x]-G^x[y])\nonumber\\
    \rightarrow  G[1]\oplus G[x]\oplus G[y]\oplus G[xy],
\end{eqnarray}
where the additional invariance under off-diagonal quadrupole moment $G[xy]$ emerges as an effect of the condensation and confinement of dual excitations.

Under periodic boundary conditions, on a $L_x\times L_y$ square lattice, the ground state degeneracy can be counted to be
\begin{eqnarray}\label{gsd_css}
    \text{GSD} = \left[ N\gcd(L_x,N) \, \gcd(L_y, N)\, \gcd(L_x, L_y, N)\right]^2,
\end{eqnarray}
 The excitations above the ground state can be either violations of $\mathcal Q_r$ or $\mathcal B_r$ plaquettes and can come into four flavors each (or combinations of these): $x$- and $y$-oriented dipolar bound states $\mathfrak{p}_x$ and $\mathfrak{p}_y$, as well as single monopoles $\mathfrak{q}$ and four-particles bound states $\mathfrak{m}$. While the three first excitation types have constrained mobility, as we discuss later in Sec. \ref{properties}, the last one is completely free to move.

Finally, we comment on the relationship between the Dipolar-Quadrupolar code in Eq.~\ref{css} and other topological orders in 2D. If we take $N=2$, the theory reduces to a $Z_2^4$ gauge theory with four copies of the toric code\cite{bulmash2018higgs}. This is evident from the stabilizer code perspective, where Eq.~\ref{css} simplifies to four independent toric codes living on four sublattice sites: $(2m,2m+1)$, $(2m,2m)$, $(2m+1,2m+1)$, and $(2m+1,2m)$. As a result, different flavors of charges (fluxes) are related by lattice translations. This echoes the fact that the anyonic excitation in most higher-rank gauge theories undergoes permutation under spatial translation, which indeed imposes mobility constraints. For a general $Z_N$ Dipolar-Quadrupolar code in Eq.~\ref{css}, the theory should resemble $Z_N$ topological order with $N^2$ types of charge (flux), featuring nontrivial mutual statistics between charges and fluxes of different types, where translation permutes anyon types\cite{pace-wen}.

\subsection{Flux attachment, Chern-Simons term, and non-CSS code}

At this point, we have established an anyon condensation process that bridges the R2TC code and the CSS version of the Dipolar-Quadrupolar code, as studied in \cite{Ebisu2023} and as schematically illustrated in Eq. \eqref{transition_delfino}, both of which are contenders for spatially modulated gauge theories in 2D with distinct charge multipole conservation laws. In this section, our goal is to generate non-CSS codes that exhibit a gauge structure similar to the model in Eq. ~\eqref{css}, recovering the models studied in Ref. \cite{delfino22}. Intriguingly, these theories can be manifested as Chern-Simons theories, where charge and flux are intrinsically linked.

To set the stage, we begin with a modified Gauss law that features charge structures similar to the ones in Eq. ~\eqref{delfino}, but with an additional constraint: unit charges are bounded to unit gauge fluxes. This flux-charge binding process has the potential to give rise to a higher-rank Chern-Simons theory \cite{you2019fractonic,delfino22} in addition to the Maxwell term.
In conventional Maxwell-type gauge theories, the ground state manifold is defined by projecting the local Hilbert space onto the vanishing net charge and flux sectors. With the flux-charge binding effect, the net charge condition will be automatically satisfied, provided the theory is flux-free. Additionally, any excitation electric charge would also contain gauge flux, and vice versa, leading to fractional statistics between charges (and fluxes).

To create a Chern-Simons-type coupling, we assign the charge density to be equal to the local flux,
\begin{align} \label{charge_flux_bind}
q_r \stackrel{!}{=}b_r,
\end{align}
where $B_r$ is defined in Eq. \eqref{css}. 
A sufficient solution to this equation is to
impose an onsite operator mapping between the two sets of $\mathbb{Z}_N$ Pauli operators $X,Z$ and $\overline X,\overline Z$, such that
\begin{align}
\bar{X}^{\dagger}_r=Z_r, ~\bar{Z}_r=X_r,
\label{localmap}
\end{align}
 In this case, the local Hilbert space per site is reduced from $N^2$ to $N$, with only one set of Pauli operators per site. The Hamiltonian in Eq.~\eqref{model1} is reduced to,
\begin{align}
H=-Y_r^{\dagger 2} \, Z_{r+e_x} \,Z_{r-e_x}\, X_{r+e_y} \,X_{r-e_y} +\text{h.c.},\label{delfino}
\end{align}
where $Y_r = Z_r X_r$.

\begin{figure}
    \centering
    \includegraphics[scale=0.3]{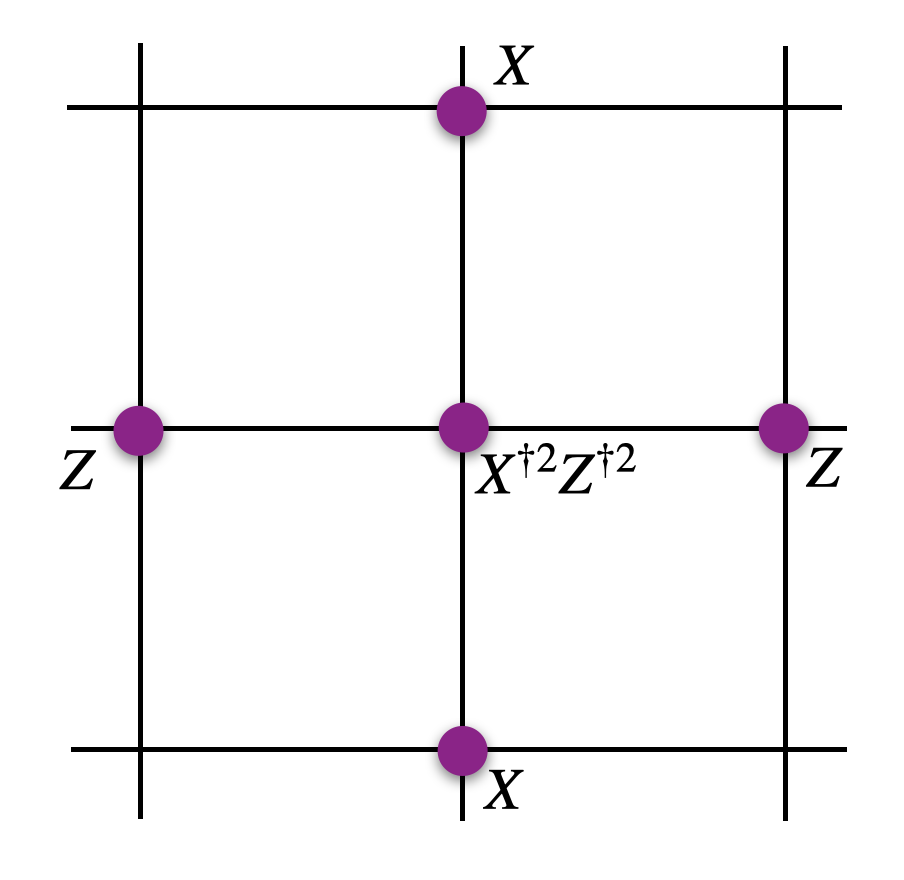}
    \caption{After the projection, the charge and flux operators are bound together, giving rise to a non-CSS code.}
    \label{Delfino-Chamon}
\end{figure}

It is worth noting that the solution given in Eq.~\eqref{localmap} is not unique. It is possible that there are other solutions that satisfy the flux-charge binding constraint, but we will defer investigation on this to future research.

From a continuum point of view, such a flux attachment process engenders a dipole Chern-Simons-like coupling
\begin{align}
 &\frac{1}{2\pi}[A_0(-\partial^2_x A^{yy}+\partial^2_y A^{xx})-A^{xx}\partial_t A^{yy}],
 \label{frac}
\end{align}
between the $A^{xx}$ and $A^{yy}$ gauge field components.
This is the theory studied in detail in Ref. \cite{delfino22} and defines a dipole-moment, and off-diagonal quadrupole moment, conserving $\mathbb Z_N$ Chern-Simons-like theory.

The theory in Eq. \eqref{frac} presents ground state degeneracy that depends on the system size and is given by
\begin{eqnarray}
    \text{GSD} = N \gcd(L_x, N)\, \gcd(L_y,N) \, \gcd(L_x, L_y, N),
\end{eqnarray}
which is very similar to the CSS code expression in Eq. \eqref{gsd_css}, but without the overall power of two. This is due precisely to the reduction in total number of charge and flux global sectors under the attachment in Eq. \eqref{charge_flux_bind}. While the doubled theory in Eq. \eqref{model1} is time reversal (TR) invariant, as expected coming from a Higgsed Maxwell-like theory, the non-CSS theory in Eq. \eqref{delfino} is not. This follows from the requirement that charge and flux, which transform differently under TR, should be bound together, and is expected from a Chern-Simons-like theory.

\subsection{Route to Moessner Code}\label{sec:mos}

To develop a comprehensive anyon condensation web that links a variety of modulated models,
we once again initiate with the R2TC and employ an alternative anyon condensation scheme that confines the diagonal components $E^{xx},E^{yy}$. Through this process, we obtain a new, exactly solvable model dubbed the \textit{Moessner code}, as its gauge structure bears similarity to the classical spin model proposed by Moessner in Ref.~\cite{benton2021topological}.

We still come up with the R2TC exploited in Eq. \eqref{r2tc}, wherein the operator $e^{i A^{xx}}\, (e^{i A^{yy}})$ generates a pair of dipole moments for both $q^x\, (q^y)$. Consider we condense these dipole moments by adding a strong Higgsing term, $\cos (A^{xx}+A^{yy})$, which imposes the constraint $(A^{xx}+A^{yy})=0\, (\text{mod } 2\pi)$. That necessarily precludes operators such as $E^{xx},E^{yy}$, that do not commute with the constraint, and as a result the vector charge $(q^x,q^y)$ in Eq. \eqref{r2tc} ceases to be well-defined. A new Gauss law, that commutes with the constraint, arises
\begin{align}
  \Delta_x q^x- \Delta_y q^y=(\Delta^2_x -\Delta^2_y )E^{xy}+\Delta_x \Delta_y (E^{xx}-E^{yy}).
\end{align}
Accordingly, the magnetic flux operator becomes
\begin{align}
& b = (\Delta^2_x -\Delta^2_y) A^{xx}-\Delta_x \Delta_y A^{xy}
\end{align}

\begin{figure}
    \centering
    \includegraphics[scale=0.31]{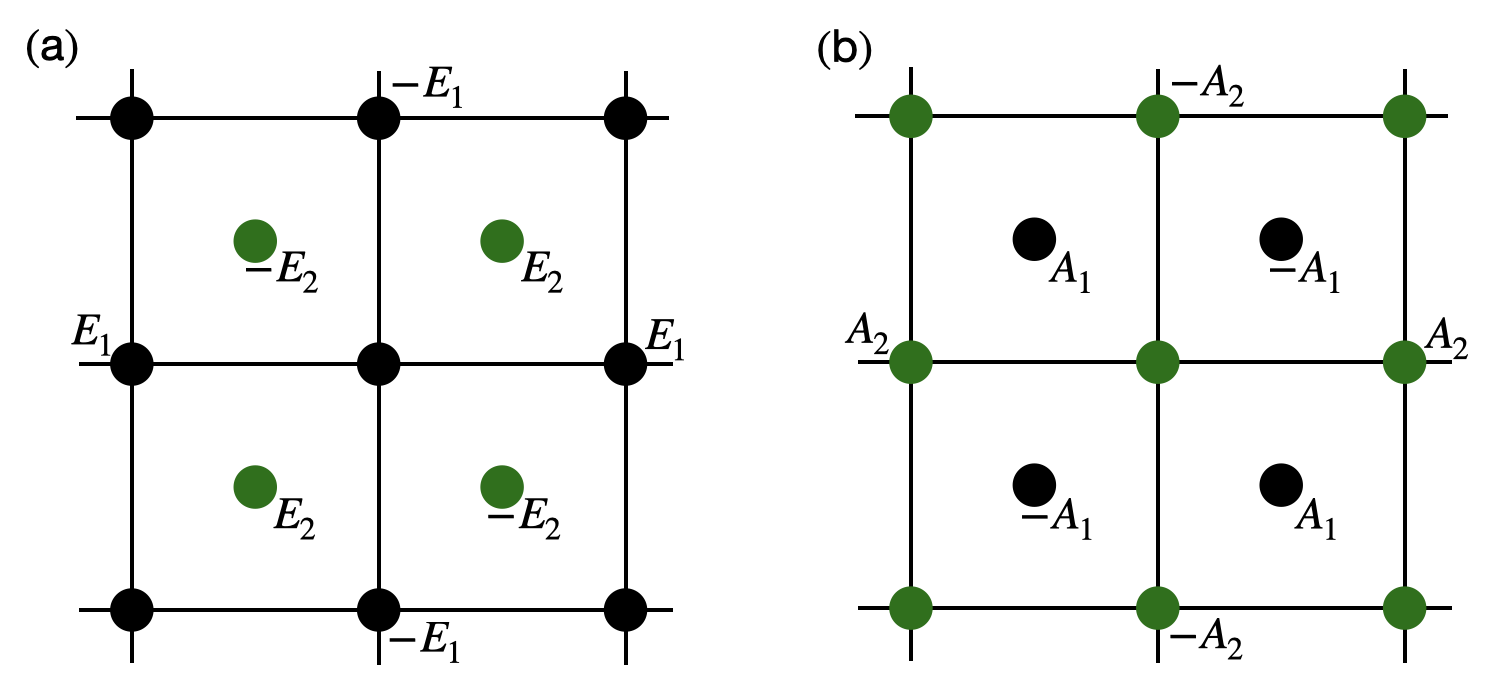}
    \caption{(a) Gauss law and (b) magnetic flux for Moessner model.}
    \label{Moessner}
\end{figure}

To avoid confusion and simplify notation, we redefine the field as,
\begin{align}
 &E^{xy} \rightarrow E^{x},~ (E^{xx}-E^{yy}) \rightarrow E^{y}\nonumber\\
  &A^{xx} \rightarrow A^{y} ,~ A^{xy} \rightarrow A^{x},
\end{align} 
where the elements $E^y,A^y$ reside at centers of plaquettes and $E^x,A^x$ occupy vertex sites.
With this, the theory's Gauss law and magnetic flux are expressed as
\begin{align}
& q=(\Delta^2_x -\Delta^2_y )E^{x}+\Delta_x \Delta_y E^{y} \nonumber\\
 & b = (\Delta^2_x -\Delta^2_y) A^{y}-\Delta_x \Delta_y A^{x}.
 \label{eq:mos}
\end{align}
The gauge potential components transform under gauge transformations as,
\begin{align}
 &A^{x} \rightarrow A^{x} +(\Delta^2_x -\Delta^2_y )f, \nonumber\\
 &A^{y} \rightarrow A^{y} +\Delta_x \Delta_y f.
\end{align}

We can also express the gauge theory in Eq.\ref{eq:mos} in terms of stabilizer code by parameterizing the field components as $e^{i\frac{2\pi}{N}E^{x}}=X, e^{i A^{x}}=Z, e^{i \frac{2\pi}{N} E^{y}}=\bar{X},$ and $e^{iA^{y}}=\bar{Z}$. The resulting CSS-type code is given by
\begin{eqnarray}\label{modelmoessner}
    H=-\sum_r\mathcal{Q}_r - \sum_r \mathcal{B}_r+\text{h.c.} 
\end{eqnarray}
with $\mathcal{Q}_r$ and $\mathcal{B}_r$ being the charge and flux operators, defined as 
\begin{widetext}
\begin{eqnarray}
\mathcal{Q}_r  &=& X^{\dagger}_{r-e_x} \, X^{\dagger}_{r+e_x} \, X_{r+e_y} \, X_{r-e_y} \,  \bar{X}^{\dagger}_{r+e_v} \, \bar{X}_{r-e_v} \, \bar{X}_{r+e_u} \, \bar{X}^{\dagger}_{r-e_u}, \nonumber\\ \text{and}\quad \mathcal{B}_r  &=& \bar{Z}^{\dagger}_{r-e_x} \, \bar{Z}^{\dagger}_{r+e_x} \, \bar{Z}_{r+e_y} \, \bar{Z}_{r-e_y} \,   Z^{\dagger}_{r-e_v} \, Z_{r+e_v} \, Z_{r-e_u} \, Z^{\dagger}_{r+e_u}, 
\label{moscode}
\end{eqnarray}
\end{widetext}
Here, $X,Z$ are situated on the vertex sites, while $\bar{Z},\bar{X}$ are located at the center of the plaquette. The vectors $e_u,e_v$ connect coordinates of vertices and centers of plaquettes, where $e_u=({e_x+e_y})/{2}$ and $ e_v=({e_x-e_y})/{2}$. 

The exactly solvable Hamiltonian in Eq. \eqref{model1} can be thought of as coming from  gauging symmetries associated with the 
dilaton transformation generator $G[x^2+y^2]$, as well as charge $G[1]$ and dipole momenta $G[x]$ and $G[y]$. Thus, this particular scheme of anyon condensation gives rise to a transition
\begin{eqnarray}\label{transition_moessner}
    G^x[1]\oplus G^y[1]\oplus (G^y[x]-G^x[y])\nonumber\\
    \rightarrow  G[1]\oplus G[x]\oplus G[y]\oplus G[x^2+y^2],
\end{eqnarray}
where we again observe the emergence of extra spatially modulated symmetries in the condensed phase.

\subsection{Modulated-Gauge principle: confinement, restricted mobility, and conservation laws}\label{mod}

We expect the protocol illustrated in Sec. \ref{sec:mos} and Sec. \ref{sec:chamon} to establish a connection among a variety of 2D modulated gauge theories through anyon condensation mechanisms, laying the groundwork for a network of spatially modulated stabilizer codes. A prevailing question is: Is there a unique correspondence between the modulated phases before and after a specific type of anyon condensation? Furthermore, can we generalize this scheme to obtain other 2D models \cite{delfino20232d, watanabe} via anyon condensation of R2TC?

Before addressing the central question, let's compare the symmetry and conservation laws between R2TC and the Dipolar-Quadrupolar code, as detailed in Sec.~\ref{sec:chamon}. R2TC embodies a gauge theory with flux and flux-dipoles
\begin{eqnarray}
    \tilde G[1] &=& \sum_r b_r\nonumber\\
\tilde G[x] &=& \sum_r x\,  b_r\nonumber\\
 \tilde G[y] &=& \sum_r y\, b_r
\end{eqnarray}
 conservation. In contrast, the Dipolar-Quadrupolar code incorporates an additional flux-quadrupole
 \begin{eqnarray}
     \tilde G[xy]  = \sum_r xy\, b_r
 \end{eqnarray}
 conservation. Implementing anyon condensation through Higgs mechanism and freezing $A_{xy}$ has the effect of confining $e^{i \frac{2\pi}{N} E^{xy}}$, responsible for transporting a flux $x$-dipole in the $y$-direction (or a flux $y$-dipole  in the $x$-direction). We review the mobility of flux and charge-bound states in Sec. \ref{properties}. This confinement results in two outcomes:

1)  Flux-dipoles experience enhanced mobility constraints, such that their motion is strictly limited to the longitudinal direction aligned with the flux-dipole orientation.

2) The $e^{i\frac{2\pi}{N} E^{xy}}$ operator, which acts by creating flux-quadrupoles, is prohibited at low energies, thereby conserving the flux-quadrupole moment $\tilde G [xy]$. 

Notably, these two outcomes are interconnected. In spatially modulated gauge theories, quasiparticle motions are restricted due to generalized conservation laws. Conversely, the conservation of charge multipole restricts the movement of charge excitations. When condensing anyons (such as charge multipoles), other excitations exhibiting non-trivial braiding with the condensed anyons (like flux multipoles) become confined. This confinement of multipole flux hinders specific kinetic movements of the flux excitations. Such quasiparticle motion constraints imply the presence of additional conservation laws, leading to a new type of modulated gauge theory.

The same ideas are behind the anyon condensation towards Moessner code in Sec.~\ref{sec:mos}. When we Higgs the theory by imposing $(A^{xx}+A^{yy})=0$, the operator $e^{i\frac{2\pi}{N}(E^{xx}+E^{yy})}$ becomes confined. This confined operator is responsible for transporting a flux $x$-dipole in the $x$-direction (or a flux $y$-dipole along the $y$-direction). Hence, the flux dipole can only move along the transverse direction perpendicular to its orientation, suggesting the conservation of flux dilaton $(x^2+y^2)B$. This conservation can be further illustrated by the fact that the monopole operator of the flux dilaton, $e^{i\frac{2\pi}{N}(E^{xx}+E^{yy})}$, is prohibited at low energy.

\begin{figure}[h!]
    \centering
    \includegraphics[scale=0.32]{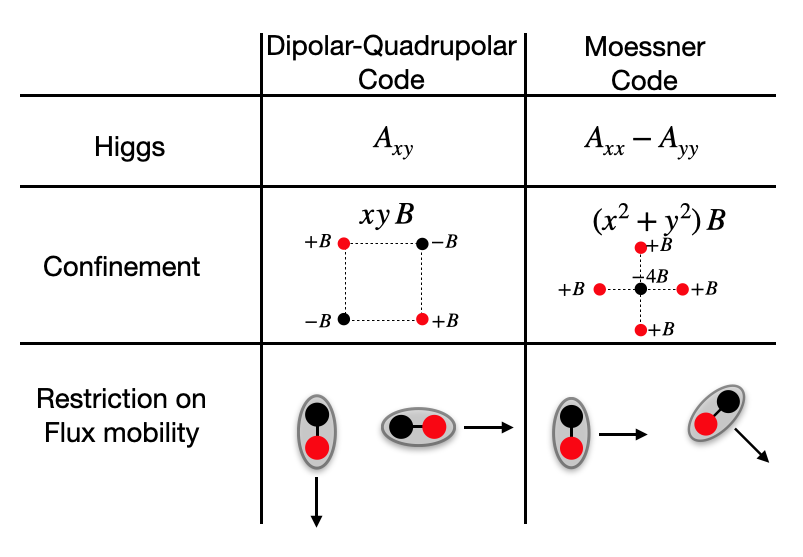}
    \caption{Higgs potential, confined bound states, and restriction on mobility of flux excitations in both Dipolar-Quadrupolar and Moessner models.}
    \label{table}
\end{figure}

In summary, the anyon condensation procedure we proposed introduces additional mobility constraints and conservation laws for higher-order charge multipoles, leading to new types of modulated gauge theories. Based on this protocol, the conserved charge multipole before anyon condensation is a subgroup of those after the condensation. This means that if we start with a higher-rank gauge theory with both charge and dipole conservation and allow various anyon condensations schemes, we can reasonably anticipate a hierarchy of spatially modulated gauge theories with charge, quadrupole, or even octupole conservation emergent laws. However, our anyon condensation protocol might not be easily generalized to phases with polynomial fractal symmetries or exponential symmetries \cite{watanabe,sala2022dynamics}. We leave this point for future exploration.

Typical anyon condensation in conventional topological field theory plays a role in \textit{reducing the topological degeneracy}\cite{Burnell_2018}. This result is intuitive: through the anyon condensation transition, some anyons become indistinguishable from the vacuum, and dual anyons (which have nontrivial statistical interactions with the condensed ones) become confined. These effects have the general role of diminishing the number of global flux sectors, consequently reducing the ground state degeneracy. In the context of gauge theories arising from gauging spatially modulated symmetries, this concept is more nuanced. In these cases, new multipole conservation laws might emerge when we condense anyons. We argue that the additional conservation laws introduce new holonomy sectors in the theory, implying that the ground state degeneracy could also increase after the condensation phase transition

A simple example of this was covered in Sec. \ref{sec:r2tc}. Starting with two copies of  standard $\mathbb Z_N$ toric codes, with a total of $N^4$ anyons, condensation of certain anyonic particles leads to R2TC whose ground state degeneracy is given by Eq. \eqref{gsd_r2tc}. For 
 for some system sizes $L_x$ and $L_y$, it can saturate its upper bound $N^6$ and effectively be more degenerate than the original uncondensed phase. One significant consideration in this example is that due to the additional conservation laws, the types of charges/fluxes of the uncondensed phase do not completely dictate the ground state degeneracy of the condensed one. Implementing anyon condensation from two copies of standard toric codes (which carry two flavors of charges, \(e_1\) and \(e_2\), and two types of flux, \(m_1\) and \(m_2\)), results in the R2TC carrying a two-component vector charge $(q^x,q^y)$ and a scalar flux $B$. Despite a qualitative decrease in the number of distinct anyon types, the additional conservation laws have the role of identifying new superselection anyon sectors coming from multi-particle bound states. As a result, the ground state degeneracy of the condensed phase increases when compared to the degeneracy of the uncondensed one.

\subsection{Implement anyon condensation by partial measurement}\label{sec:measurement}

Building on our previous discussions, we have shown that various 2D modulated gauge theories can be connected through anyon condensation. This is accomplished by introducing a strong onsite potential term so that some gauge potential components completely freeze. The resulting fluctuations to the canonical conjugated electric fields, upon perturbative expansion, yield a new type of modulated gauge theory.

In this section, we propose an alternative approach to implementing the anyon condensate, which relies on the decoherence and partial measurements of wave functions. We begin with the ground state wave function of the R2TC Hamiltonian in Eq. \eqref{r2tc}. By partially measuring some of the qubits of the wave function, the post-measurement state resembles the ground state of the Dipolar-Quadrupolar code. In this scenario, the measurement process plays a role as anyon condensation. For comparison, the essence of an anyon condensate involves enforcing an operator into its low-energy eigenstate by adding onsite interactions to the Hamiltonian. The partial measurement procedure naturally projects certain qubits of the wave function into their eigenstates.

To obtain the Dipolar-Quadrupolar code in Eq. \eqref{precha} from R2TC, we need to Higgs the off-diagonal component $A^{xy}$, as mentioned in Sec. \ref{sec:chamon}. This can be achieved by performing a measurement of the operator $\cos(A^{xy})$ on all sites. The post-measurement result would fix the value of $A^{xy}$, which is analogous to the process of anyon condensation through Higgsing $A^{xy}$. We now demonstrate that the post-measurement state is equivalent to the ground state wave function of the Dipolar-Quadrupolar code. The argument proceeds as follows. The ground state wave function of R2TC can be expressed in terms of a tensor-type wave function that encapsulates an equal weight superposition of all possible patterns of $A^{xy}_{{r}},A^{xx}_{{r}},A^{yy}_{{r}}$, with the local constraint $\Delta^2_x A^{yy}+\Delta^2_y A^{xx}-\Delta_x \Delta_y A^{xy}=0$. Suppose the post-measurement outcome on site ${r}$ is $\bar{A}^{xy}_{{r}}$, the post-measurement state then becomes a superposition of all possible patterns of $A^{xx}_{{r}}, A^{yy}_{{r}}$ that meet the local constraint
\begin{align}
 \Delta^2_x A^{yy}+\Delta^2_y A^{xx}=\Delta_x \Delta_y \bar{A}^{xy}.
\end{align}
The right-hand side of the equation is a constant, which we denote as the background flux $\Delta_x \Delta_y \bar{A}^{xy} = \bar{B}_r $. This post-measurement wave function is similar to the gauge theory we defined for the Dipolar-Quadrupolar code, with one key difference: the ground state of the Dipolar-Quadrupolar code has a net flux per site, while our result ends up with a background flux that depends on the measurement outcome of $\bar{A}^{xy}_r$. 

Our measurement protocol can also be applied to other anyon condensation schemes. As an example, one can begin with the wave function of R2TC and, upon measuring the diagonal component, the post-measurement wave function becomes akin to the ground state of the Moessner model, as described in Sec. \ref{sec:mos}. 

The intriguing and emerging effects of measurements on the evolution of quantum many-body states have sparked increased interest from both the condensed matter and quantum information communities. Recently, the potential to create exotic long-range entangled states, e.g., topological order or fracton topological ordered state, through the use of adaptive circuits has been proposed \cite{verresen2021efficiently,lu2022measurement,choi2021emergent,cotler2023emergent} had been studying extensively. While we will not delve into the effects of measurement on the 2D modulated SETs web in detail in this paper, we hope our discussion sheds light on the potential for connecting and manipulating various spatially modulated states through partial measurement.

\subsection{Advantage and limitations of our anyon condensation protocol}

To this end, we introduced a protocol for anyon condensation that facilitates the connection between various discrete higher-rank gauge theories. Anyon condensation serves as a useful mechanism for bridging different topological orders, including fracton topological orders. A comprehensive defect network approach for intertwining various topological quantum field theories (TQFT) and higher-rank gauge theories through anyon condensation at network interfaces or junctions is detailed in Ref.~\cite{aasen2020topological,song2023topological,bulmash2018higgs}.
We briefly illustrate both the limitations and advantages of our anyon condensation approach in comparison to others. As detailed in the Modulated-Gauge principle (see Sec.~\ref{mod}), a distinctive aspect of our approach is that anyon condensation not only introduces new conservation laws but also generates additional topological sectors. This increase in ground state degeneracy means that the conserved charge multipoles before anyon condensation become a subgroup of those post-condensation. Consequently, our protocol cannot connect two topological phases or higher-rank gauge theories unless one is a subgroup of the other, although most of these examples can still be connected via a defect network approach~\cite{aasen2020topological,song2023topological,bulmash2018higgs}.
Nonetheless, our protocol offers an alternative advancement: it can be implemented either through the addition of an on-site potential or by conducting a single-site measurement. This contrasts with the general protocols in defect networks, which typically require many-body interactions or simultaneous measurements of several operators across multiple sites. Notably, the on-site measurement we illustrated in Sec~\ref{sec:measurement} is applicable in Rydberg atom arrays as proposed and realized in Ref.~\cite{verresen2021efficiently,iqbal2023creation,chen2023realizing,iqbal2023topological}. This method requires only the natural atomic interactions for time evolution, followed by a single qubit measurement.

 \section{Geometry defects in SETs: an Anyon condensation view}

In this section, we explore the role of geometry defects in spatially modulated gauge theories. As explored in Refs. \cite{ you2019non,manoj2020screw,aitchison2023no}, lattice defects in fracton and modulated theories might have the role of permuting anyons, as well as introducing non-Abelian zero modes in the theory. 
Here we study geometric defects in spatially modulated gauge theory naturally realized  as branch lines subject to specific anyon condensation. 
As we show, the condensations of mobile  anyons along a branch cut play a role in creating an edge dislocation, while in a similar way, the condensation of diagonally moving bound states has the general role of creating disclination defects.
For simplicity, we study excitations in the Dipolar-Quadrupolar model, whose properties we quickly review in the next section.

\subsection{Excitations and String operators}\label{properties}

\begin{figure}
    \centering
    \includegraphics[scale=0.3]{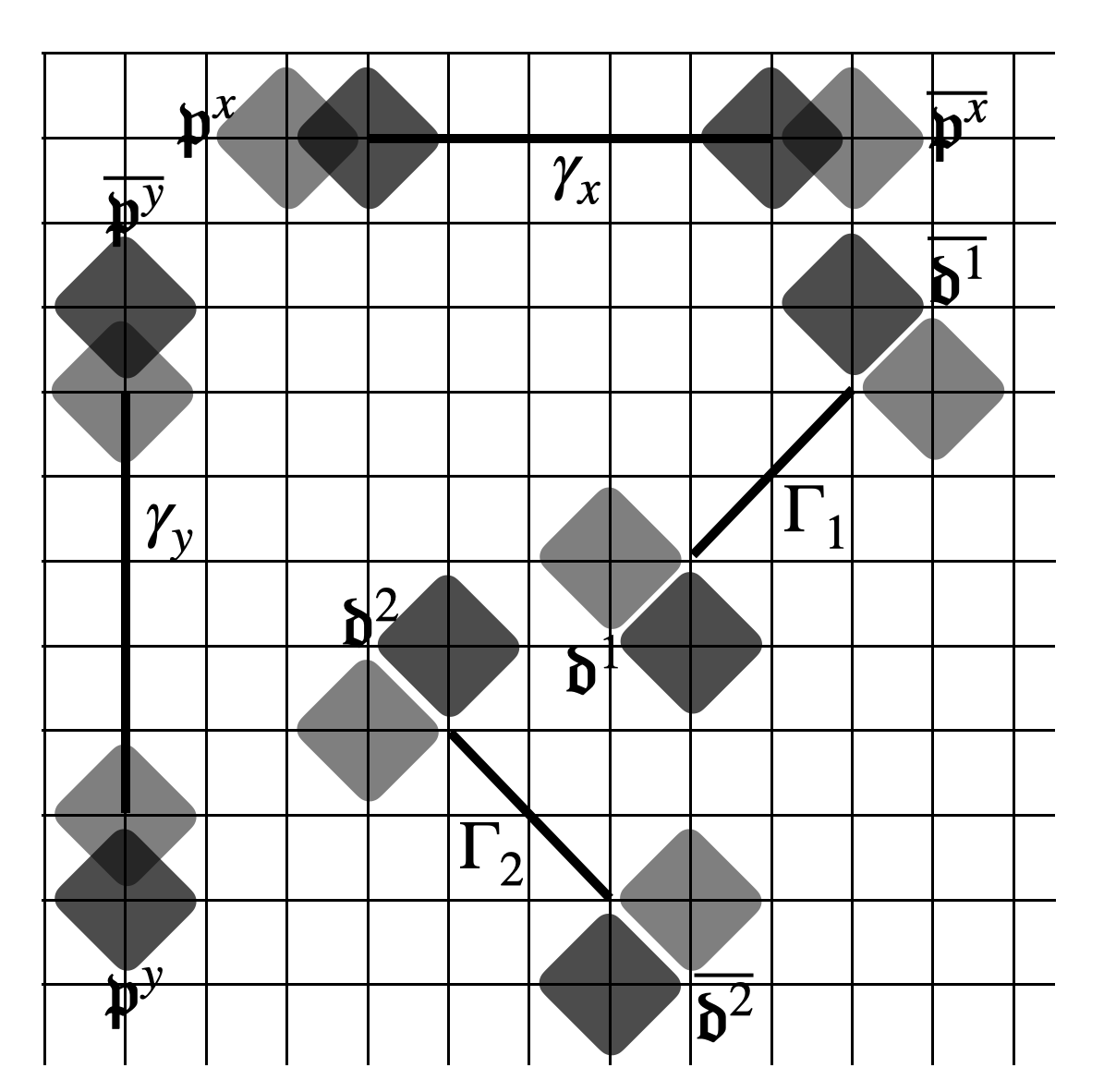}
    \caption{Anyons $\mathfrak p$ and $\mathfrak{d}$ and their corresponding string operators with support on $\gamma_x$, $\gamma_y$, $\Gamma_1$, and $\Gamma_2$}
    \label{fig:strings_pd}
\end{figure}

Let us quickly review the allowed symmetry respecting dynamics of the excitations of Dipolar-Quadrupolar code in Eq. \eqref{delfino}. There are four types of excitations: monopole particles $\mathfrak{q}$, dipolar bound states $\mathfrak{p}_{x}$ and $\mathfrak p_y$, and four-particle bound states $\mathfrak{m}$ anyons. The $\mathfrak{m}$ particles are completely mobile and can be created at the endpoints of the completely flexible string
\begin{eqnarray}\label{strings_anyons}
	&\mathfrak{m}&:\begin{cases}
 W(\gamma_x) = \prod_{\gamma_x} X_r X_{r+e_y}^\dagger,\\
 W(\gamma_y) = \prod_{\gamma_y} Z_r Z_{r+e_x}^\dagger,\end{cases}
\end{eqnarray}
where $\gamma_x$ and $\gamma_y$ are $x$- and $y$-oriented string sections. The dipolar bound states $\mathfrak{p}^{x}$ and $\mathfrak p^y$ can only move along straight lines, in a lineon-like behavior, and are created at the endpoints of the string operators
\begin{eqnarray}
	&\mathfrak{p}^x&: V(\gamma_x) = \prod_{\gamma_x} X_r \nonumber\\
	&\mathfrak{p}^y&: V(\gamma_y) = \prod_{\gamma_y} Z_r,
\end{eqnarray}
as illustrated in Fig. \ref{fig:strings_pd}.
Lastly, the isolated excitations can only be moved through fixed-length $N$ string operators
\begin{eqnarray}
    &\mathfrak{q}&: \begin{cases}
	    U(\lambda_x) = \prod_{\lambda_x} X_r^x, \\
     U(\lambda_y) = \prod_{\lambda_y}, Z_r^y\end{cases}
\end{eqnarray}
for straight strings $\lambda_x$ and $\lambda_y$ whose size are $N$ (or integer multiples of $N$).

Additionally, there are also ``diagonal''-dipoles $\mathfrak{d}^1$ and $\mathfrak{d}^2$, created at the endpoint of diagonal strings $\Gamma_1$ and $\Gamma_2$
\begin{eqnarray}
	S(\Gamma_1) &=&  \prod_{\Gamma_1} X_r Z_r\nonumber\\
	S(\Gamma_2)&=& 	 \prod_{\Gamma_2} X_r Z_r^\dagger,
\end{eqnarray}
as illustrated in Fig. \ref{fig:strings_pd}.
These dipolar bound states are not completely independent from the $\mathfrak p^x$ and $\mathfrak p^y$ ones. Instead, they are the result of fusion
\begin{eqnarray}
	\mathfrak{d}^1 = \mathfrak{p}^x\times \mathfrak{p}^y \quad \text{and} \quad 	\mathfrak{d}^2 = \mathfrak{p}^x\times \overline{\mathfrak{p}^y}.
\end{eqnarray}
 We note that, similarly to the $V(\gamma_x)$ and $V(\gamma_y)$ strings, the $S(\Gamma_1) $ and $S(\Gamma_2)$ ones have support along straight lines $\Gamma_1$ and $\Gamma_2$, which and cannot bend. Otherwise, additional excitations are created around the region of the bent region.

\subsection{Anyon Condensate: $\mathfrak{m}$-dipole}

\begin{figure}[h!]
    \centering
    \includegraphics[scale=0.3]{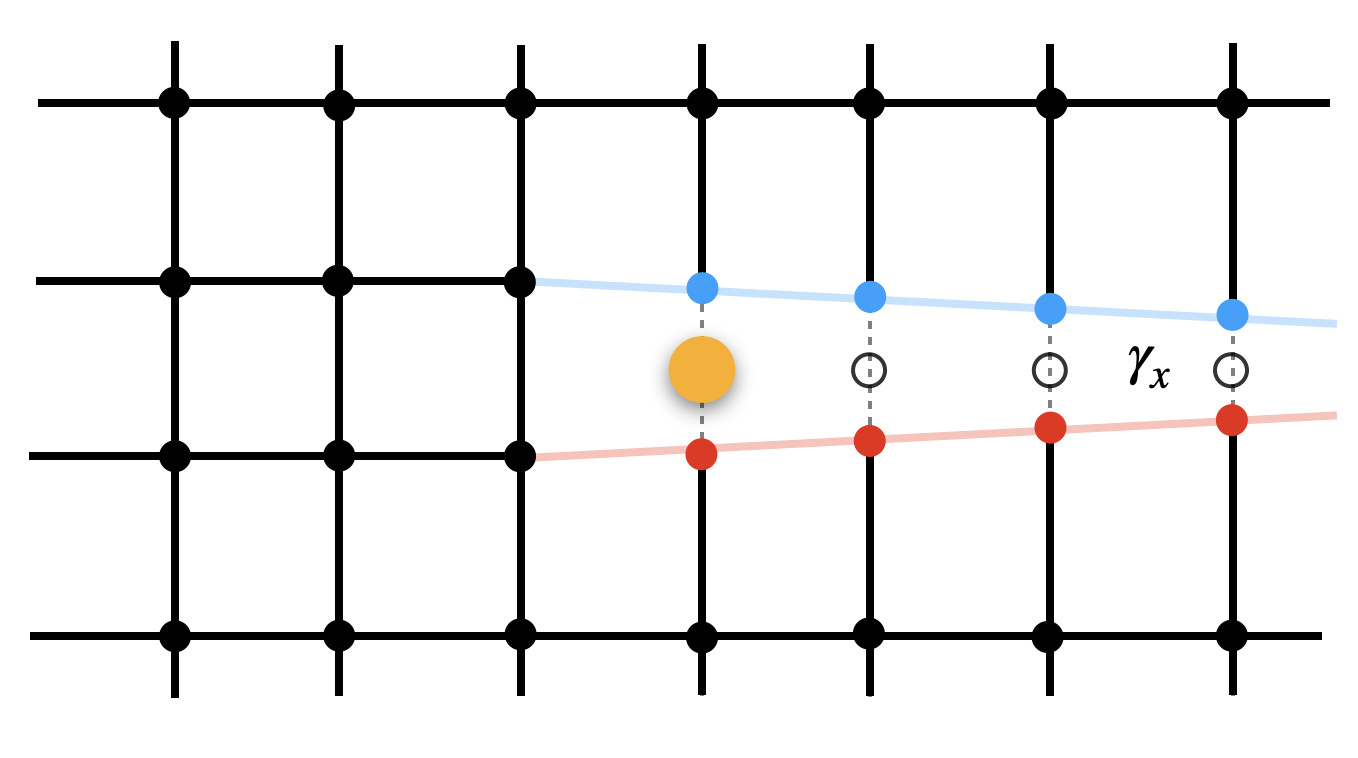}
    \caption{Condensing $\mathfrak m $ anyons along a branch $\gamma_x$ effectively implements a lattice dislocation defect.}
    \label{dislocation}
\end{figure}

Let us start by spatially condensing \( \mathfrak{m} \) dipoles along a defect line $\gamma_x$ and demonstrate that such anyon condensation on the defect branch is akin to adding a dislocation defect. To begin with, we apply a strong Ising coupling along the \( y \) link \( X_r X_{r+e_y}^\dagger \), which is responsible for condensing a dipole consisting of a pair of \( \mathfrak{m} \) anyons
\begin{eqnarray}
	H \rightarrow H - g\sum_{r\in \gamma_x} X_r X_{r+e_y}^\dagger.
\end{eqnarray}
The condensation along the branch is illustrated in Fig. \eqref{dislocation}.

In the strong $g$ coupling limit, we consider only states in the Hilbert space that obey $X_r X_{r+e_y}^\dagger \ket{\psi} = +1\ket\psi$. We then use a perturbative expansion on the stabilizers near the branch to obtain an effective Hamiltonian. For this, let us explicitly write
\begin{eqnarray}
	H = g\left (H_0+\dfrac{1}{g} H_1\right ),
\end{eqnarray}
where $H_1 = H_{DQ}$ is the Dipolar-Quadrupolar Hamiltonian in Eq. \eqref{delfino}. The perturbative parameter is $g^{-1}$ and $H_0 = \sum_{r\in \gamma_x} X_r X_{r+e_y}^\dagger + \text{h.c.}$ is the strong field applied along $\gamma_x$ 

\begin{figure}
\centering  	
\includegraphics[scale=0.45]{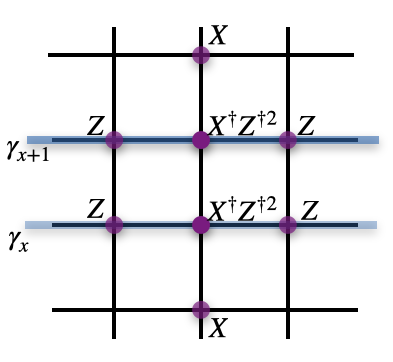}
\caption{New plaquette operators $\tilde B_{r}$, in second-order perturbation theory,  overlapping with the double string at $\gamma_x$ and $\gamma_{x+1}$, generated in second order in perturbation theory.}
\label{plaq}
\end{figure}

Let \( \mathcal P \) be a projection operator onto the states satisfying the constraint \( X_r X_{r+e_y}^\dagger = +1 \) for all \( x \) in \( \gamma_x \) and let $B_r$ abbreviate each one of the terms in Hamiltonian \eqref{delfino}. Then, in first-order perturbation theory, the terms in the Hamiltonian that overlap with the condensed branch $\gamma_x$ vanish. The effective Hamiltonian, in first-order
\begin{eqnarray}
	H^{(1)}_{\text{eff}} = \mathcal P H_1 \mathcal P =  - \sum_{r \notin \gamma_x} B_{ r} + \text{h.c.},
\end{eqnarray}
only contain terms that are not in the string $\gamma_x$. This follow because the plaquette terms $B_{{r}}$ do not commute with the constraint $X_r X_{r+e_y}^\dagger = +1$ for $r \in \gamma_x$. In second order, however, there are non-trivial contributions to the effective Hamiltonian near the branch
\begin{eqnarray}
	H^{(2)}_{\text{eff}} &= & \mathcal P H_1 \dfrac{(1-\mathcal P)}{E-H_0}H_1\mathcal P \nonumber\\
	&\ni& - K \sum_{r\in \gamma_x} \tilde B_{r} + \text{h.c.}
 \label{eff}
\end{eqnarray}
where $K\sim \mathcal{O}(g^{-2})$ and $\tilde B_{ r}$ for $r \in \gamma_x$ is depicted in Fig. \ref{plaq}.

Upon condensing the \( \mathfrak{m} \)-dipole along the branch cut by projecting \( X_r X_{r+e_y}^\dagger = +1 \), the two qubits at sites \( x \) and \( x+e_y \) become constrained, and can be effectively treated as a single degree of freedom located at an intermediate site. In other words, the dipole condensation introduces a dislocation where two adjacent rows along the $x$-direction shrink into a single row, as depicted in Fig \ref{dislocation}. The effective perturbative Hamiltonian we derived in Eq. \eqref{eff} indeed produces the stabilizer Hamiltonian of the Dipolar-Quadrupolar code on a dislocation lattice as Fig.~\ref{dislocation}.

When we condense the \( \mathfrak{m} \)-dipole along the branch cut, reminiscent of creating a translation defect (dislocation) that merges two adjacent $x$-rows, both the total charge and \( x \)-dipole remain conserved quantities, as
\begin{eqnarray}
    \prod_{r} B_{r} = \mathds{1} \quad \text{and}\quad      \prod_{r} B^{\rho_x x}_{r} = \mathds{1},
\end{eqnarray}
still hold.
However, since the dislocation literally mixes the positions between \( r \) and \( r+e_y \) along the branch cut, the \( y \)-dipole and \( xy \)-quadrupole moment of $B_r$ become ill-defined when passing through the branch cut. In particular, if we count the total number of \( y \)-dipoles and \( xy \)-quadrupole moments in the presence of dislocation lines,
\begin{eqnarray}
    \prod_{r} B^{\rho_y y}_{r} &=& W(\gamma_{x+1})^{\rho_y}\nonumber\\
  \prod_{r} B^{\rho_{xy} x y}_{r} &=& W(\gamma_{x+1})^{\rho_{xy} \,x},\label{no_constraint}
\end{eqnarray}
where $W(\gamma_x)$ is the string-like operator defined in Eq. \eqref{strings_anyons}. 
The fact that Eq. \eqref{no_constraint} does not act as an identity in the entire Hilbert space under closed boundary conditions implies that the $y$-dipole and $xy$-quadrupole momenta are no longer conserved quantities and as a consequence $\mathfrak{p}^y$ and $\mathfrak{m}$ no longer correspond to well defined anyonic superselection sectors in the theory.

\subsection{Anyon Condensate: $\mathfrak d^1$-particle}

\begin{figure}[h!]
    \centering
    \includegraphics[scale=0.3]{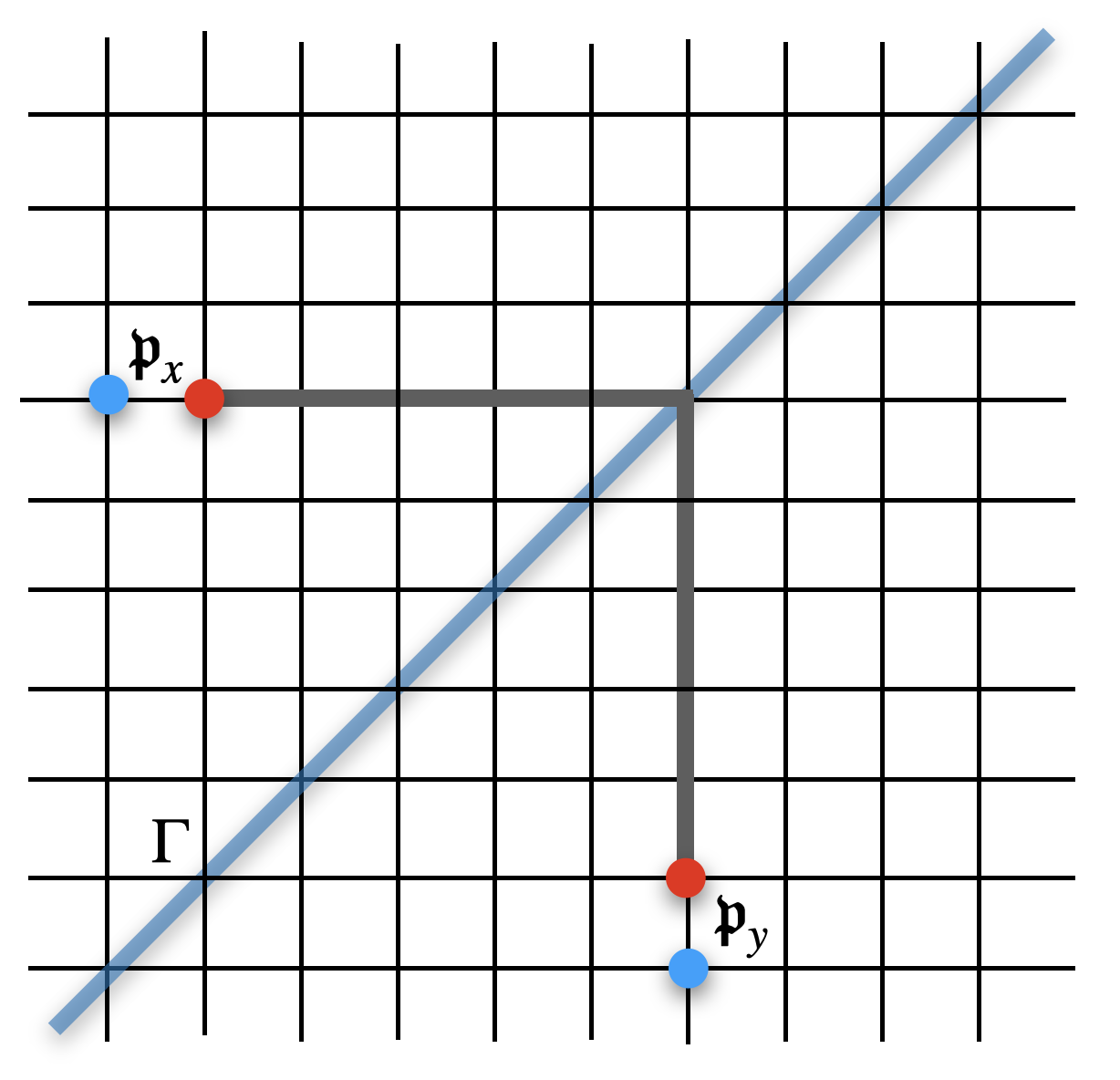}
    \caption{The presence of a $\mathfrak{d}^1$ condensate along the line $\Gamma$ breaks the total $x$ and $y$ dipole conservation and allows $\mathfrak p^x $ and $\mathfrak p^y$ anyons to convert into each other.}
    \label{px_py}
\end{figure}

\begin{figure}
    \centering
    \includegraphics[scale=.4]{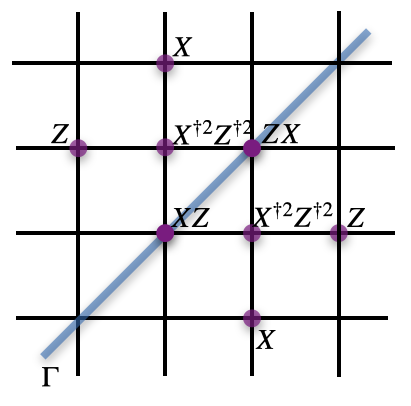}
    \caption{New plaquette operators $ B'_{r}$ in second order perturbation theory.}
    \label{plaq_diag}
\end{figure}

Condensates of \( \mathfrak d^1 \) anyons can offer a more intriguing scenario. Similar to the protocol we elucidated in the previous section, we can introduce a strong onsite field 
\begin{eqnarray}
    H\rightarrow H - h\sum_{\Gamma_1} (Z_rX_r+h.c.)
\end{eqnarray} 
acting along a diagonal line $\Gamma_1$ (Fig. \ref{px_py})  that proliferates $ \mathfrak d^1 $ anyons. Under a perturbative expansion, the effective plaquette stabilizers \( B'_{r} \) around the branch cut, which commutes with the $ \mathfrak d^1 $ condensate, are shown in Fig. \ref{plaq_diag}. The effect of the defect line is to transmute $ \mathfrak p^x $ anyons into $\mathfrak p^y$ anyons as they cross the condensation line, as showed in Fig. \ref{px_py}.

If we create a $ \mathfrak p^x $ excitation along the $x$-direction and cross the defect branch cut, the resulting excitation becomes $ \mathfrak p^y $ anyons, which can only move along the $y$-direction. In this regard, the role of the $ \mathfrak d^1 $-particle condensate along the branch cut is reminiscent of a disclination line (a rotational defect) that can permute two types of anyons—$\mathfrak p^x$ and $ \mathfrak p^y $—which display nontrivial braiding statistics with different subdimensional mobilities. What sets its peculiar character is that the $ \mathfrak d^1 $ anyon condensation along the $\Gamma$ introduces additional non-Abelian defects at the endpoints of the line, as referenced in Ref.~\cite{You2013_synthetic,you2019non}. We plan to address this matter in a future study.

\section{Multipartite Entanglement mutual information}

\subsection{Holonomies and Wilson algebra}

In this section, we derive the Wilson algebra of the  code, as presented in Eq. \eqref{delfino}. A detailed derivation of the Wilson line operator can be found in the Appendix; here, we only summarize the results. There are four types of Wilson operators along the $y$-direction,
\begin{align}
W_1(x) & =  \prod_y X(x,y) X^{\dagger}(x+1,y) , \nonumber\\  
W_2(x) & =  \prod_y X^{y}(x,y) X^{\dagger y}(x+1,y), \nonumber\\ 
W_3(x) & =  \prod_y X(x,y) , ~W_4(x)  = \prod_y X^{y}(x,y).
\label{eq:wil}
\end{align}

The Wilson operators are subjected to the uniform condition: $\partial^2_x W_3 (x) =\partial^2_x W_4 (x) =0$. This indicates that we need to pin the value of two nearby Wilson lines to establish the value of all holonomies generated by $W_3,W_4$. Further, once we set the values of $W_3$ and $W_4$, other Wilson lines like $W_1,W_2$ can be uniquely determined, since $\partial_x W_3=W_1,\partial_x W_4=W_2$. This implies the existence of four independent Wilson operators along the $y$-direction. We can apply the same logic to identify the dual Wilson operators along the $x$-lines.
\begin{align}
V_1(y) & =  \prod_x Z(x,y) Z^{\dagger}(x,y+1) , \nonumber\\  
V_2(y) & =  \prod_x Z^{x}(x,y) Z^{\dagger x}(x,y+1) ,\nonumber\\ 
V_3(y) & =  \prod_x Z(x,y), ~V_4(y)  =  \prod_x Z^{x}(x,y).
\label{eq:wil2}
\end{align}
 It is worth mentioning that all the $W_i$ and $V_i$ operators have support on closed single/double parallel strings. 
Not all $W_i$ and $V_i$ operators commute with each other, as they have a non-vanishing intersection. The non-trivial algebra between these operators and its action on the Hilbert space produces the degeneracy of the ground state space (as well as a topological degeneracy in all other energy sectors).

In order to specify the Wilson line algebra and recover the ground state degeneracy, we identify the set of Wilson line operators that generate the ground state manifold. From our previous demonstration, it is not hard to conclude that the eigenvalue of operators $V_i(0), W_i(0)$ for $i=1,\ldots,4$ fix the ground state Hilbert space so we can use them to span ground state manifold, 
\begin{align}
V_3(0) W_3(0) & = e^{-i\frac{2\pi}{N}} W_3(0) V_3(0) \nonumber\\  
V_1(0) W_4(0) & = e^{i\frac{2\pi}{N}} W_4(0) V_1(0) ,\nonumber\\ 
V_4(0) W_1(0) & = e^{i\frac{2\pi }{N}} W_4(0) V_1(0), \nonumber\\
V_2(0) W_2(0) & = e^{i\frac{2\pi}{N}} W_2(0) V_2(0).
\end{align}
Now we comment on the ground state degeneracy. The operators $V_3(y),W_3(x)$ are $\mathbb{Z}_N$ operators, regardless of the system size. It is not difficult to show that these operators span an $N$-fold degenerate Hilbert space.
Extra caution is needed when evaluating the eigenvalues of the remaining operator pairs. For $W_4(x)$, it reduces to a $\mathbb{Z}_{\gcd(L_y,N)}$ operator under closed boundary conditions, meaning its eigenvalues can only change mod ${\gcd(L_y,N)}$. The same reasoning applies to $V_4(y)$ and $W_1(x)$, which together span a $\gcd(L_x,N)$-fold degenerate Hilbert space.
Finally, consider the pair $V_2(y), W_2(x)$. Under closed boundary conditions, $V_2(y)$ reduces to a $Z_{\gcd(L_x,N)}$ operator and its eigenvalues can only change mod ${\gcd(L_x,N)}$. Similarly, $W_2(x)$ reduces to a $Z_{\gcd(L_y,N)}$ operator and its eigenvalues can only change mod ${\gcd(L_y,N)}$. Thus, $V_2(y), W_2(x)$ span a $\gcd(L_x,L_y, N)$-fold degenerate Hilbert space.

\subsection{Diagnosing Modulated Behavior Using Mutual Information}

Many intriguing aspects of modulated gauge theories stem from UV/IR mixing \cite{you2019emergent,xu2007bond,paramekanti2002ring,tay2010possible,seiberg2020exotic}, where the behavior at low energy can be influenced by the specific details of the lattice. This phenomenon is compelling because it seemingly contradicts the principles of topological quantum field theories, where low-energy physics arises from topologically robust patterns due to entanglement \cite{haah2014bifurcation,ma2018topological,he2018entanglement,shirley2019universal} and often remains unaffected by UV properties  \cite{levin-wen,levin2005string}. When it comes to modulated gauge theories, a pertinent question is whether one can visualize UV-IR mixing through observable quantities, such as correlation functions or entanglement mutual information. In this section, we show that the emergence of UV-IR mixing in spatially modulated gauge theories can be detected through entanglement entropy and mutual information. More specifically, we will evaluate the long-range mutual information between distant rows of qubits under various geometric cuts, which sheds light on the inherent correlation between different Wilson line operators.

In the referenced work~\cite{jian2015long,liu2023multipartite}, the authors demonstrate that both topological order and symmetry-breaking states can be detected and diagnosed using long-range mutual information (LRMI). This formalism is based on the fact that a topological ground state, when on a closed manifold, exhibits long-range correlations between non-local string operators. Starting with a maximally entangled state on a half-torus, the Wilson line operators—defined on the open cylinder—emerge as cat states exhibiting maximal uncertainty. This state resembles the cat state of the 1D quantum Ising model, except that in the topologically ordered state, the order parameter assumes the form of a non-contractible string. Thus, it is natural to expect a non-vanishing long-range mutual information (LRMI) between two distant non-contractible regions (such as stripes), which symbolize string-like order parameters circling the non-contractible loops. The existence of non-vanishing mutual information between stripes implies that the Wilson line operator defined on different stripe regions shares the same eigenvalue, so determining the pattern in one stripe would subsequently reduce the information entropy of the other stripe. In this section, our aim is to adapt and expand upon this concept for 2D modulated codes, such as the Dipolar-Quadrupolar code. Intriguingly, due to the UV-IR mixing nature inherent in modulated gauge theories, we witness that the entanglement LRMI is acutely sensitive to both the geometry and distance of the cut.

\begin{figure}[h!]
    \centering
    \includegraphics[scale=0.3]{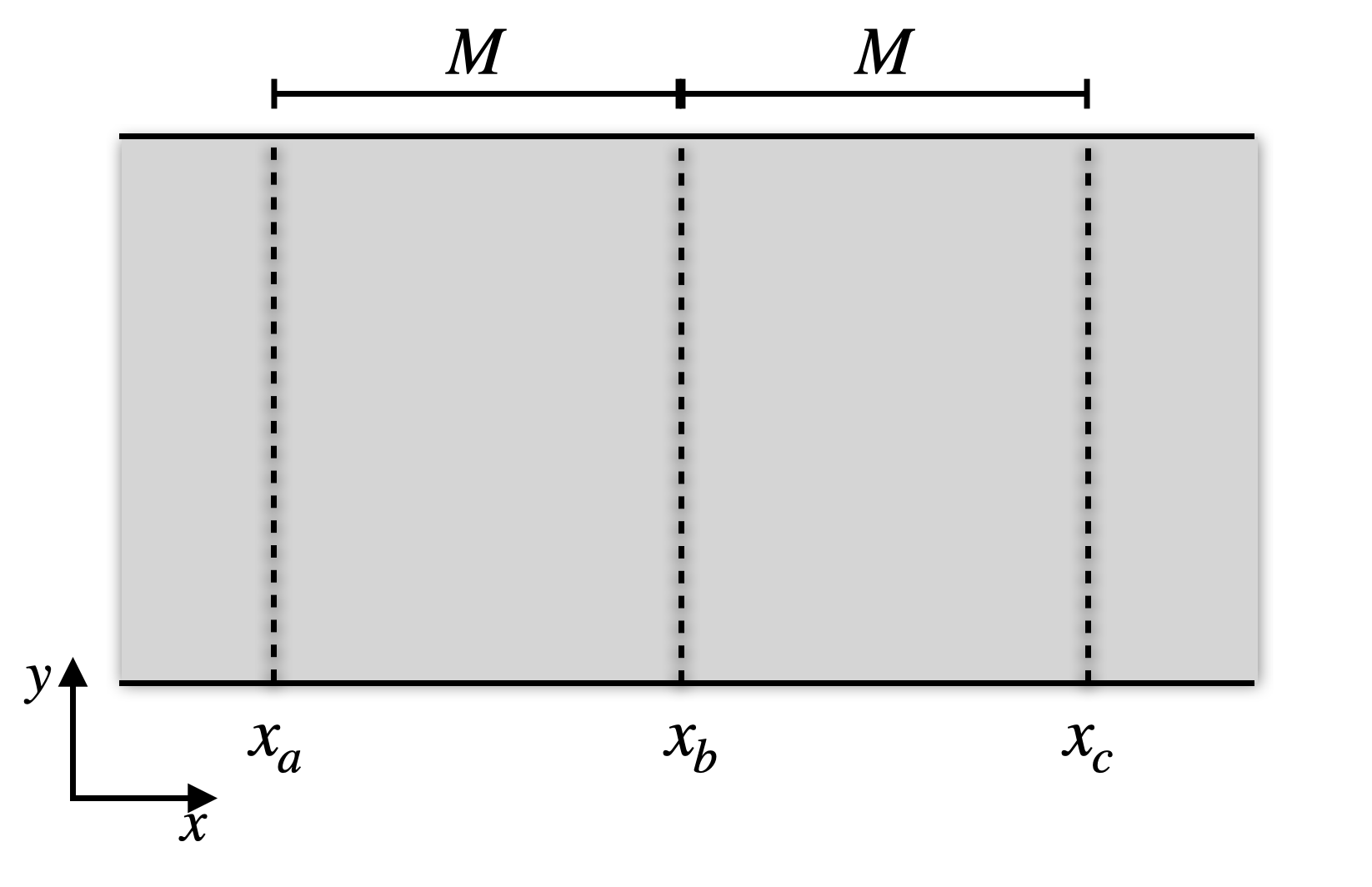}
    \caption{We compute the tripartite mutual information among the three rows illustrated as dashed lines at $x_a$, $x_b$, and $x_c$.}
    \label{mutual_info}
\end{figure}

We focus on the Dipolar-Quadrupolar code in Eq.~\ref{delfino} as an example. 
Suppose we take out three rows of qubits along the $y$-direction and label their position as $x_a, x_b, x_c$ (with $x_a-x_b=b_c-x_c=M$) as Fig.~\ref{mutual_info}. The mutual information among these three rows indicates the information entropy shared over large distances. Let us assume we start with a specific ground state, which is the eigenstate of all Wilson line operators $V_i$ along the $x$-direction. In other words, all $W_i$ operators along the $y$-direction appear as cat states with maximal uncertainty. We now calculate the tripartite mutual information among these three regions.
\begin{align}
    I(a:b:c)=S(a\cup b \cup c)+S(a)+S(b)+S(c)\nonumber\\
    -S(c\cup b )-S(a\cup c )-S(a\cup b )
\end{align}
Here $a,b,c$ refer to the three rows at $x_a, x_b, x_c$.
The entanglement entropy $S$ offsets the information entropy contributed by locally fluctuating patterns in each region, and the residual LRMI $I(a:b:c)$ documents the long-range correlations between them.

We first analyze the entropy produced by each row at $x_i$. Given that the ground state wave function is projected by the stabilizer operators in the Hamiltonian, the entanglement entropy of a single row or sets of rows depends on two factors: 1) The number of independent stabilizers, including combinations of several stabilizers, that act directly on the row or sets of rows, and 2) The constraints on certain global operators, like Wilson line operators, that are independent of local stabilizers.

Let us start with $S(a)$, which quantifies the information entropy of the qubits on row $x_a$. There are no stabilizers, nor combinations thereof, that act on the row of $x_a$ independently. Additionally, given that the $W_i$ operators along the $y$-direction appear as cat states with maximal uncertainty, there are no global constraints on the row of $x_a$. Consequently, $S(a)$ has only local contributions with its entropy being $L_y \ln(N)$. The same principle applies to $S(b)$ and $S(c)$.

Now, let's examine $S(c\cup b)$. There are no stabilizers, nor combinations thereof, that act on the rows on $x_b \cup x_c$ independently. Given that all $W_i$ operators along the $y$-direction manifest as cat states with maximal uncertainty, there appears to be no global constraint on the two distant rows at $x_a,x_b$. Nevertheless, the conditions $\partial^2_x W_3=0$ and $\partial_x W^2_4=0$ necessitate that (see appendix for full derivations),
\begin{align}
(W_3(x_a) W^{-1}_3(x_b))^{\frac{L_x}{\gcd(L_x,M)}}=1\nonumber\\
 (W_4(x_a) W^{-1}_4(x_b))^{\frac{L_x}{\gcd(L_x,M)}}=1
 \label{res}
\end{align}
The operator $W_3(x_a) W^{-1}_3(x_b)$ is a $Z_N$ operator, but the constraint from Eq.~\ref{res} reduces its eigenvalue to a $Z_{\gcd(\frac{L_x}{\gcd(L_x,M)},N)}$ value. Likewise, $W_4(x_a) W^{-1}_4(x_b)$ is a $Z_{\gcd(L_y,N)}$ operator, but the constraint reduces its eigenvalue to a $Z_{\gcd(\frac{L_x}{\gcd(L_x,M)},N,L_y)}$ value.
This reduces the information entropy of the two rows by
\begin{align}
\ln(N)+\ln(\gcd(L_y,N))-\ln\left( \gcd\left(\frac{L_x}{\gcd(L_x,M)},N\right)\right)\nonumber\\
-\ln\left( \gcd\left(\frac{L_x}{\gcd(L_x,M)},N,L_y\right)\right)
\end{align}
in addition to the entropy contributed by local fluctuations. By applying the same logic, the information entropy of $S(a\cup b )$ is reduced by
\begin{align}
\ln(N)+\ln(\gcd(L_y,N))-\ln\left( \gcd\left(\frac{L_x}{\gcd(L_x,M)},N\right)\right)\nonumber\\
-\ln\left( \gcd\left(\frac{L_x}{\gcd(L_x,M)},N,L_y\right)\right)
\end{align}
while that of $S(a\cup c)$ is reduced by \begin{align}
\ln(N)+\ln(\gcd(L_y,N))-\ln\left( \gcd\left(\frac{L_x}{\gcd(L_x,2M)},N\right)\right)\nonumber\\
-\ln\left( \gcd\left(\frac{L_x}{\gcd(L_x,2M)},N,L_y\right)\right)
\end{align}

Now we look into $S(c\cup b\cup a )$. As long as we know the patterns on the rows at $x_a,x_b$, the Wilson line operators $W_3(x_a)$ and $W_4(x_a)$ are determined, leading to a reduction in total information entropy by $\ln(N)+\ln(\gcd(L_y,N))$. Additionally, the restrictions in Eq.~\ref{res} further decrease the information entropy by \begin{align}
\ln(N)+\ln(\gcd(L_y,N))-\ln\left( \gcd\left(\frac{L_x}{\gcd(L_x,M)},N\right)\right)\nonumber\\
-\ln\left( \gcd\left(\frac{L_x}{\gcd(L_x,M)},N,L_y\right)\right)
\end{align}
Based on these arguments, it can be concluded that the tripartite mutual information takes the following form:
\begin{widetext}
\begin{align}
I(a:b:c)=\ln(N)+\ln(\gcd(L_y,N))
   -\ln\left( \gcd\left(\frac{L_x}{\gcd(L_x,M)},N\right)\right)
-\ln\left( \gcd\left(\frac{L_x}{\gcd(L_x,M)},N,L_y\right)\right)\nonumber\\
      -\ln\left( \gcd\left(\frac{L_x}{\gcd(L_x,2M)},N\right)\right)
-\ln\left( \gcd\left(\frac{L_x}{\gcd(L_x,2M)},N,L_y\right)\right)
    \label{lrmi}
\end{align}
\end{widetext}
The LRMI depends on both the system size and the distance between the three areas, which is distinctly different from the mutual information in conventional gauge theories\cite{jian2015long}. More precisely, the LRMI of modulated gauge theories is influenced by the system size as well as the distance between the cuts. By altering the distance M between the a-b-c region, the LRMI exhibits periodic fluctuations. This starkly contrasts with LRMI in conventional gauge theory, where mutual information saturates to a constant at long distances.

The novelty of modulated gauge theories arises from the fact that the Wilson operators must adhere to a specific geometric pattern without local deformation. Consequently, if we calculate the tri-partite mutual information between distant rows along different directions by varying cuts, the LRMI can change dramatically depending on the geometric cut. To substantiate this statement, we rotate the system by $\pi/4$ and examine the qubits along the rows in the $\hat{x}\pm\hat{y}$ directions. Here we redefined the coordinate as:
 \begin{align}
     \sqrt{2} r_1=x+y,   \sqrt{2} r_2=x-y,
 \end{align}

 The magnetic flux operator becomes,
 \begin{align}
     &(\partial^2_{r_1} +\partial^2_{r_2})(A_{yy}-A_{xx})+ \partial_{r_2}\partial_{r_1} (A_{xx}+A_{yy})=B,\nonumber\\
     &A_{xx}-A_{yy}\rightarrow A_{xx}-A_{yy}+\partial_{r_2}\partial_{r_1}\alpha \nonumber\\
     &A_{xx}+A_{yy}\rightarrow A_{xx}+A_{yy}+(\partial^2_{r_1} +\partial^2_{r_2}) \alpha
 \end{align}

If we remove a single row along the $r_1$ direction, the only non-contractible operator that commutes with the Hamiltonian is:
 \begin{align}
     &\int (A_{xx}-A_{yy}) \, dr_1=U(r_2), ~\partial^2_{r_2} U=0
 \end{align}
which can be treated as a Wilson line operator along the $r_1$ direction.
We now place the theory under periodic boundary conditions along $r_1,r_2$ and choose the ground state as a cat state for which both the $U$ and $\nabla_{r_2}U$ operators reach maximal uncertainty. We then remove three rows of qubits along the $r_1$ direction and label the columns as $r_a,r_b,r_c$ ($r_a-r_b=r_b-r_c=2M$).
Based on our previous argument, one can conclude that tripartite mutual information takes the following form:
\begin{align}
 I(a:b:c)=\ln(N) -\ln\left( \gcd\left(\frac{L_2}{\gcd(L_2,M)},N\right)\right)\nonumber\\
-\ln\left( \gcd\left(\frac{L_2}{\gcd(L_2,2M)},N\right)\right)
\end{align}
$L_2$ is the number of sites along $r_2$ direction. Upon comparing this result with Eq. \eqref{lrmi}, it becomes evident that the long-range mutual information (LRMI) between three distant regions is influenced not only by their distance but also by the orientation of the cut, thereby introducing a level of geometric dependence. This sensitivity to geometry can be interpreted through the perspective of the Wilson operators, which might adhere to a specific geometric shape without the flexibility to bend. Since LRMI is governed by the correlations among the Wilson operators, changing the direction or geometry of the cut also changes the number of Wilson line operators that survive within the cut, which results in a change in LRMI.

\section{Outlook}

In this work, we have introduced an anyon condensation framework that links different two-dimensional spatially modulated gauge theories. As we explicitly studied through some examples, a general effect of the condensate is the emergence of additional higher-multipole momenta conservations, which directly affects the quasi-particle content, as well as their allowed dynamics.

For future perusal, we highlight a few open questions that are worth exploring in the future: 1) As discussed in Sec.~\ref{sec:measurement}, anyon condensation can be manipulated by making partial measurements upon the wavefunction. Since measurements can be achieved by introducing decoherence channels in an open system, it is feasible that the anyon condensation scheme we have introduced in this manuscript could be realized through quantum decoherence. This scenario extends the exploration of spatially modulated states in open systems and provides a feasible platform for building quantum memories in Noisy Intermediate-Scale Quantum (NISQ) devices.
2) Likewise, since the geometric defect introduces additional anyon condensation defects to the modulated gauge theory, we expect that impurities and lattice defects can engender a zoology of new exotic spatially modulated states. 3) We studied a multi-partite mutual information protocol that is able to detect UV/IR mixing information in the ground state wavefunction, in contrast to the usual topological entanglement topological entropy. We expect that it can applied to various 3D fracton theories, opening a new chapter in the exploration of novel entanglement features in higher-dimensional fracton phases

\emph{Acknowledgement}--- 
We are grateful to Claudio Chamon for enlightening discussions and valuable comments in the draft. 
 This work was completed in part at Aspen Center for Physics (Y.Y.) and at Paths to Quantum Field Theory 2023 workshop at Durham University (G.D.). It is supported by National Science Foundation grant PHY-2210452 and Durand Fund (Y.Y) as well as the DOE Grant No. DE-FG02-06ER46316 (G. D.).

\section*{Appendix}
\label{appendix}

\subsection*{Wilson Algebra for Dipolar-Quadrupolar code}\label{global_constraints}
Historically, the holonomies engendered by the Wilson line operators manifest the global flux sectors to which the ground state on a torus belongs. Building on this line of thinking, we show how to obtain Wilson operators pertinent to the Dipolar-Quadrupolar code from the underlying gauge theory. For higher-rank gauge theories, the `Wilson operators' creating immobile quasi-particle excitations turn out to be richer and more diverse than in the conventional $\mathbb{Z}_N$ gauge theory for the following reasons. 1) Due to the restricted mobility of the quasiparticles, some of the Wilson lines need to be straight and geometrically oriented in a specific direction. 2) There might exist other `Wilson operators' defined on a non-contractible manifold, such as membrane, cage, or fractal, that are responsible for the holonomies of higher-rank gauge theory~\cite{haah2011local,you2020fractonic,prem2019cage}. 3) Different Wilson operators that are parallel to each other may not render the same value, as opposed to the conventional $\mathbb{Z}_N$ gauge theory whose Wilson line operators are invariant under translation. For higher rank gauge theory, the dipole and quadruple moments transform non-trivially under translation, and so does the global flux sector. Consequently, two parallel flux lines might return different values.

Recall that in the usual 2D $\mathbb{Z}_N$ gauge theory, the magnetic flux is given by $B = \partial_x A_y - \partial_y A_x$ and the total flux on the half cylinder ${\cal A}$ bounded by at $x=x_0$ and $x=x_n$ is be characterized by parallel Wilson line operators
\begin{align} \int_{\cal A} B \,dV= \oint A_y(x_n , y ) dy-\oint A_y(x_0 , y ) dy=0 ,  \nonumber \end{align} 
With the integral $\oint$ going around the full circumference of the cylinder. The net flux condition ($\int B dV = 0$) implies that the two parallel Wilson lines render the same value. Since the two Wilson lines are spatially separated while the Hamiltonian is local, each $\oint A_y(x , y) dy$ must commute with all local terms in the Hamiltonian and can be treated as a global flux operator that characterizes the holonomy. One obtains another Wilson line operator along the $y$-direction from the charge sector, i.e. $\oint E_y (x, y ) dy$. These two comprise all possible Wilson lines along the $y$-loop.

\begin{figure}[h!]
    \centering
    \includegraphics[scale=0.30]{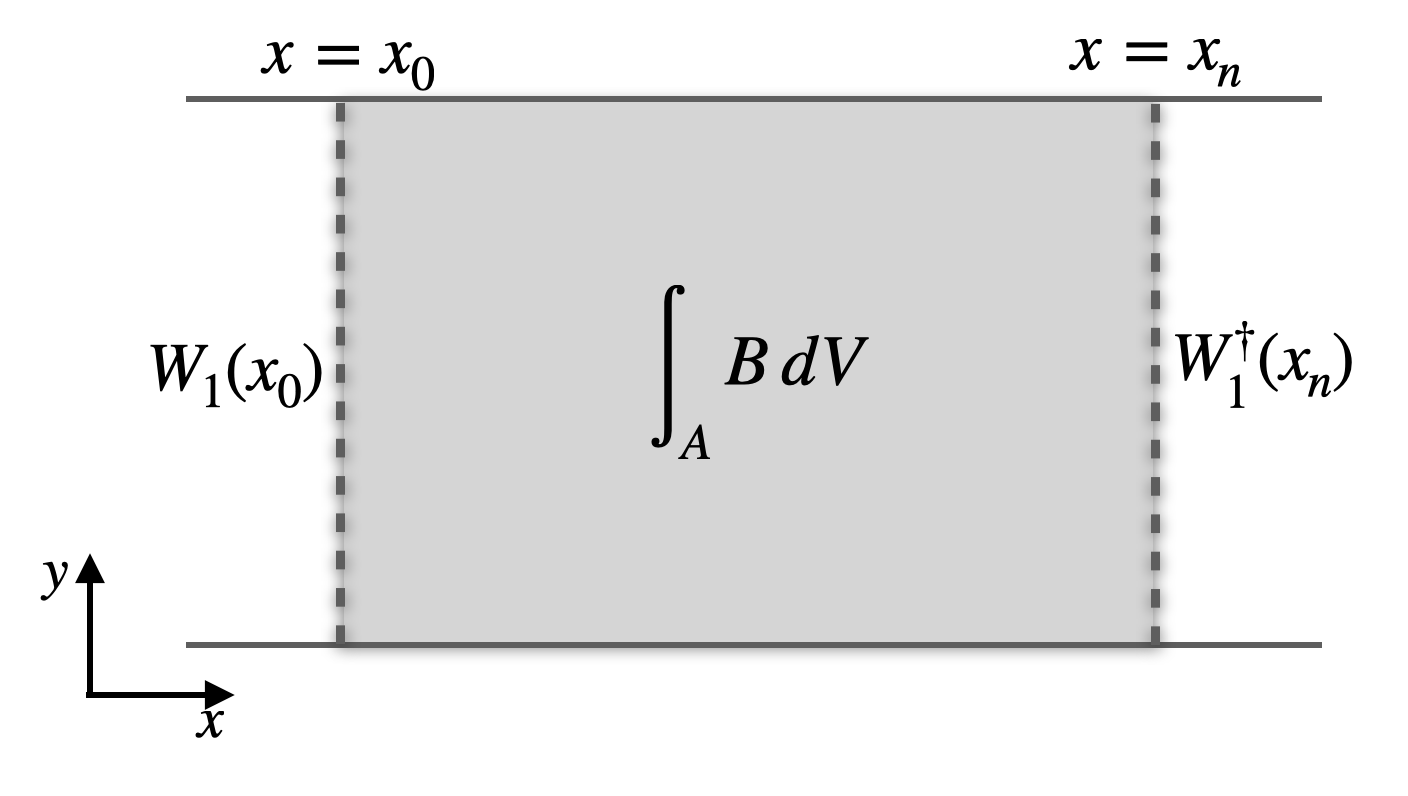}
    \caption{The integral of conserved quantities in a finite region $\mathcal A$ results into Wilson lines at the boundary $\partial \mathcal A$ of the region $\mathcal A$.}
    \label{region}
\end{figure}

In this appendix, we derive the Wilson operators of the non-CSS version of the Dipolar-Quadrupolar code, as represented in Eq. \eqref{delfino}. We begin with the definition of the flux operator $b=\partial^2_x A^{yy}-\partial^2_y A^{xx}$. Given that the Dipolar-Quadrupolar code can be characterized by a Chern-Simons type gauge theory, the pattern of the ground state on a closed manifold is based on the net flux condition $b=0$. The magnetic charges represented by $b$ demonstrate a number of conservation laws associated to $\tilde G[1], \tilde G[x],\tilde G[y],\tilde G[xy]$.

Suppose we place the ground state wave function on an open cylinder and focus on the Wilson lines defined along the $y$-loop. In this scenario, we can study the holonomies associated to $\tilde G[1]$ and $\tilde G[x]$ by integrating them in a finite region $\mathcal A$, as showed in Fig. \ref{region} 
\begin{align}
&\prod_{r\in \mathcal A} \mathcal B_r =  \prod_{y} e^{i \Delta_x A^{yy}(x_n , y )}e^{-i \Delta_x A^{yy}(x_0 , y)},\nonumber\\
&\prod_{r\in \mathcal A} \mathcal B_r^y = \prod_{y} e^{i \Delta_x yA^{yy}(x_n , y)} e^{-i\Delta_x yA^{yy}(x_0 , y)}  .  \label{eq:qxqy-conserve-2}
\end{align}
where $\Delta_x$ refers to the lattice difference along the $x$-direction. Based on Eq.~\eqref{eq:qxqy-conserve-2}, the total flux on an open cylinder is reduced to two operators localized on the boundary. Following the notation in Sec.~\ref{sec:chamon}, we express the gauge potential in terms of Pauli operators $X=e^{i A^{yy}},Z=e^{i A^{xx}}$ and obtain the Wilson operators,
\begin{align}
W_1(x) & =  \prod_y X(x,y) X^{\dagger}(x+1,y) , \nonumber\\  
W_2(x) & =  \prod_y X^{y}(x,y) X^{\dagger y}(x+1,y)
\end{align}
Here we choose a coordinate that the site (x,y) resides on the lattice characterized by integer coordinates.
Due to the flux conservation law, Eq. (\ref{eq:qxqy-conserve-2}), these two operators are uniform along the $x$-coordinate: $\Delta_x W_2 (x) =\Delta_x W_1 (x) =0$.

Similarly, the other two conserved quantities $\tilde G[y] $ and $\tilde G[xy]$ engender another set of Wilson line operators,
\begin{align}
W_3(x) & =  \prod_y X(x,y) ,\nonumber\\
W_4(x) & = \prod_y X^{y}(x,y).
\end{align}
Due to the flux conservation law, they are nonuniform along the $x$-coordinate: $\Delta^2_x W_3 (x) =\Delta^2_x W_4 (x) =0$. This indicates the necessity of pinning the value of two proximate Wilson lines to establish the value of all holonomies generated by $W_1,W_2$. However, once the values of the aforementioned Wilson operators $W_1,W_2$ are fixed, $W_3(x),W_4(x)$ can be uniquely determined, since $\Delta_x W_3=W_1,\Delta_x W_4=W_2$. This implies the existence of four independent Wilson operators along the $y$-direction. The same argument can be applied to identify the dual Wilson operators along the $x$-lines.

\begin{align}
V_1(y) & =  \prod_x Z(x,y) Z^{\dagger}(x,y+1) , \nonumber\\  
V_2(y) & =  \prod_x Z^{x}(x,y) Z^{\dagger x}(x,y+1) ,\nonumber\\ 
V_3(y) & =  \prod_x Z(x,y), \nonumber\\
V_4(y) & =  \prod_x Z^{x}(x,y).
\end{align}

\subsection*{Information entropy for Wilson line operators}\label{information}

In this appendix, we provide a detailed derivation for the constraint of the Wilson line operator in Eq.~\eqref{res}.
 There are no stabilizers, nor combinations thereof, that act on the rows on $x_b \cup x_c$ independently. Nevertheless, the conditions $\partial^2_x W_3=0$ and $\partial_x W^2_4=0$ indicate the value of $(W_3(x_a) W^{-1}_3(x_b))$ and $(W_4(x_a) W^{-1}_4(x_b))$ are uniform under translation. In the presence of periodic boundary conditions, multiply these operators along $x$-direction ${\frac{L_x}{\gcd(L_x,M)}}$ times gives unity:
 
\begin{align}
 (W_3(x_a) W^{-1}_3(x_b))^{\frac{L_x}{\gcd(L_x,M)}}=1\nonumber\\
 (W_4(x_a) W^{-1}_4(x_b))^{\frac{L_x}{\gcd(L_x,M)}}=1
\end{align}
The operator $W_3(x_a) W^{-1}_3(x_b)$, by definition, is a $Z_N$ operator, but the constraint reduces its eigenvalue to a $\gcd(\frac{L_x}{\gcd(L_x,M)},N)$ value. Likewise, $W_4(x_a) W^{-1}_4(x_b)$ is a $Z_{\gcd(L_y,N)}$ operator, but the constraint reduces its eigenvalue to a $\gcd\left(\frac{L_x}{\gcd(L_x,M)},N,L_y\right)$ value.

\bibliography{biblioanyon.bib}

\end{document}